\documentclass[aps,twocolumn,prx,floatfix]{revtex4-2}

\usepackage{xcolor,hyperref}
\hypersetup{
   colorlinks,
   linkcolor={blue!50!black},
   citecolor={blue!50!black},
   urlcolor={blue!80!black}
}

\usepackage{epsfig,amsmath}
\usepackage{bm}
\usepackage{soul,ulem}
\usepackage{hyperref}
\usepackage{footmisc}
\usepackage{wrapfig}
\DeclareMathAlphabet{\mathitbf}{OML}{cmm}{b}{it}

\renewcommand{\=}{\!=\!}

\renewcommand{\figurename}{Fig.}

\DeclareMathAlphabet\mathbfcal{OMS}{cmsy}{b}{n}
\begin{document}

\title{Lab earthquakes confirm the theory of frictional slip pulses}
\author{Alina Shafir$^1$}
\email{Contributed equally}
\author{Tom Gabrieli$^2$}
\email{Contributed equally}
\author{Yuval Tal$^2$}
\author{Eran Bouchbinder$^1$}
\email{eran.bouchbinder@weizmann.ac.il}
\affiliation{$^1$Chemical and Biological Physics Department, Weizmann Institute of Science, Rehovot 7610001, Israel\\
$^2$Department of Earth and Environmental Sciences, Ben-Gurion University of the Negev, Beer-Sheva 84105, Israel}

\begin{abstract}
Large natural earthquakes are typically mediated by frictional pulse-like rupture, which features a finite slipping zone.
Recently, a comprehensive two-dimensional theory of frictional slip pulses has been developed. It predicts that pulses are categorically unstable rupture modes, whose evolution is intrinsically slow. Unsteady pulses satisfy an equation of motion expressed in terms of discrete observables, which is inherently related to their steady-state counterparts. The theory also predicts a transition from decaying to slowly growing pulses. Here, we perform extensive lab earthquake experiments to test the theory. The experiments confirm the theoretical predictions for pulse-like lab earthquakes over a range of prestress levels, rupture nucleation conditions and small-scale fault roughness amplitudes. Specifically, the predicted time-dependent dynamics are tested in a plane defined by the evolving pulse size and peak slip rate, along with measurements of the dimensionless growth rate of pulses, demonstrating their intrinsically slow unsteady nature. The predicted transition between decaying and growing pulses is also experimentally demonstrated. These results constitute major progress in understanding a dominant earthquake rupture mode.
\end{abstract}

\maketitle

\section{I\lowercase{ntroduction}}

The spatiotemporal dynamics of frictional systems, such as natural faults and the contacting tectonic plates that form them, are mediated by the propagation of interfacial ruptures~\cite{Ben-Zion2001,Scholz2002,Lu2007,Lu2010a,Svetlizky2019}. Once triggered, rupture propagation weakens (unlocks) the contact interface and frictional strength reduction is typically accompanied by rapid sliding that releases large amounts of previously stored elastic energy. On Earth, such rupture-driven energy release generates an earthquake, which radiates seismic waves that can produce severe ground motion and associated hazards~\cite{Scholz2002}.

Interfacial failure is classified into two distinct categories~\cite{Ben-Zion2001,Scholz2002,Lu2007,Lu2010a,Svetlizky2019}, one corresponds to expanding crack-like rupture and the other to pulse-like rupture, commonly termed slip pulses~\cite{freund1979mechanics,Heaton1990,Perrin1995,Beroza_Mikumo_1996,Beeler1996,Cochard1996,Andrews1997,Zheng1998,Nielsen2000,Nielsen2003,Brener2005,Dunham2005,lykotrafitis2006self,Shi2008,Rubin2009,Garagash2012,platt2015steadily,Gabriel2012,Putelat2017,Michel2017,kohrangi2019pulse,chen2020cascading,wu2023pulse,ROCH2022104607,galetzka2015slip,Dunham2011a,Brener2018,pomyalov2023self,brantut2019stability,pomyalov2023dynamics,pomyalov2024,gabrieli2025lab,bouchbinder2026equation}. Crack-like rupture features a slipping zone and slip duration at an interfacial point that increase with the rupture propagation distance, while in pulses they are independent of it. That is, slip pulses are compact objects that feature a characteristic size. Understanding and predicting the spatiotemporal dynamics of frictional rupture is of prime importance in diverse fields, including geophysics, materials science, mechanics and tribology.

Slip pulses constitute a dominant rupture mode in natural earthquakes and likely so in many manmade engineering systems. Yet, their basic understanding is not as developed as that of their crack-like counterparts. Much of our understanding of crack-like interfacial rupture is based on analogies to opening (tensile) cracks in bulk materials, where the crack faces fully detach and do not interact behind the rupture edge. The physics of slip pulses, which feature both leading and trailing edges, is qualitatively different as contact interactions at the fault can lead to a vanishing slip rate $v$ and re-strengthening/re-locking processes~\cite{Dieterich1972,Dieterich1979,Ruina1983,Marone1998a,Nakatani2001,Baumberger2006Solid,Ben-David2010}, entirely absent in bulk failure.

These frictional re-strengthening processes select the finite size $L$ of pulses, which are commonly termed `self-healing' pulses (pulses can also emerge due to geometrical effects~\cite{weng2019dynamics} and material contrast across the fault~\cite{Shlomai2016,poles2024slip}, not considered here). Only recently, a comprehensive theory of the dynamics of self-healing slip pulses has been developed~\cite{Brener2018,brantut2019stability,pomyalov2023self,pomyalov2023dynamics,pomyalov2024,bouchbinder2026equation}. Here we test the theory, to be described in more detail next, using extensive lab earthquake experiments.\\
\begin{figure*}[ht!]
    \centering
    \includegraphics[width=1.9\columnwidth]{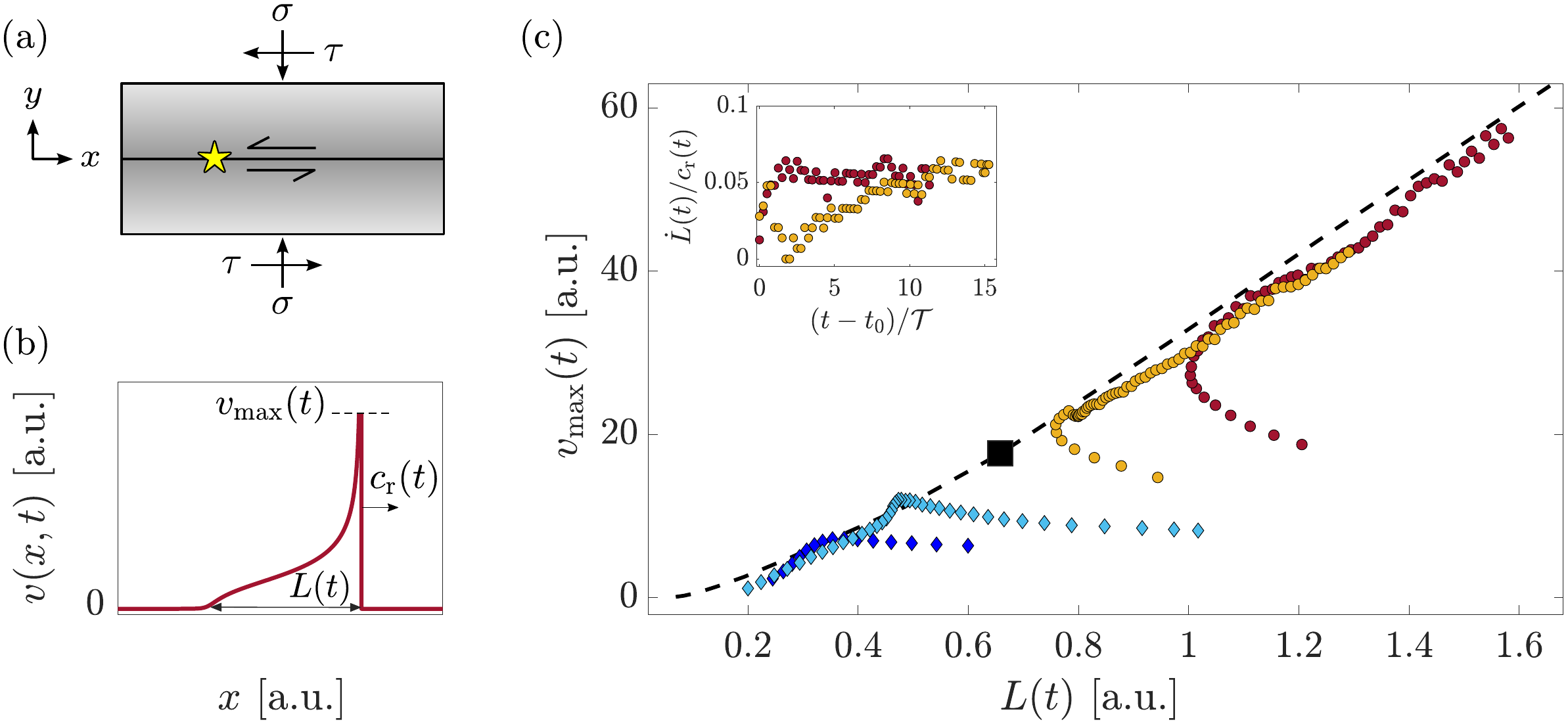}
    \caption{\footnotesize{\bf Theoretical predictions for the dynamics of slip pulses.} (a) A sketch of a frictional system, composed of two large and identical linear elastic blocks, forming a nominally flat contact interface (fault). The coordinate along the fault is $x$ and it is located at $y\!=\!0$, where $y$ is a fault-perpendicular coordinate. The third dimension of the blocks is either small compared to the two other dimensions or the system's dynamics are independent of it, such that the system is effectively two dimensional (2D). The fault prestress (background stress) features a normal compressive component $\sigma$ and a shear component $\tau$. Rupture is nucleated at a localized region along the fault (the hypocenter, yellow star), giving rise to inhomogeneous sliding. (b) An example of a slip pulse solution obtained in the rate-and-state friction constitutive framework~\cite{pomyalov2023self,pomyalov2023dynamics}, where the slip rate field $v(x,t)$ is plotted at a given time $t$. The slip rate essentially vanishes both ahead of the pulse's leading edge, where $v$ rises abruptly, and behind its trailing edge, where $v$ varies smoothly. This behavior of $v(x,t)$ is accompanied by fault weakening (reduction in the frictional strength) behind the leading edge and re-strengthening (frictional strength recovery) near the trailing edge (not shown). The pulse propagates from left to right at an instantaneous speed $c_{\rm r}(t)$, featuring an instantaneous size $L(t)$ and a peak slip rate $v_{\rm max}(t)$. (c) The $v_{\mbox{\tiny \rm max}}^{(\rm ss)}[L^{(\rm ss)}]$ line (dashed line) is obtained from steady-state pulse solutions to Eq.~\eqref{eq:BIM}, coupled to rate-and-state friction under anti-plane shear (mode-III) conditions~\cite{pomyalov2023self}. The normal stress $\sigma$ is kept fixed and $\tau$ is varied, parameterizing the steady-state line. The 4 discrete datasets represent the corresponding dynamic (i.e., time dependent) pulse solutions to Eq.~\eqref{eq:BIM} at a fixed value of the shear prestress $\tau$ (whose steady-state point on the $v_{\mbox{\tiny \rm max}}^{(\rm ss)}[L^{(\rm ss)}]$ line is marked by the black square). The dynamic pulse solutions are presented in terms of the observables $L(t)$ and $v_{\rm max}(t)$ (see axis labels), i.e., each point is parameterized by $t$. The datasets differ in the rupture nucleation conditions, where the nucleation intensity decreases from the top curve to the bottom one. All unsteady pulses are dynamically attracted to the steady-state $v_{\mbox{\tiny \rm max}}^{(\rm ss)}[L^{(\rm ss)}]$ line, and once meeting it (at a time point to be denoted by $t_0$), they subsequently adhere to it, as predicted by Eq.~\eqref{eq:eom}. Hot-colored circles (two upper datasets) correspond to growing pulses and cold-colored diamonds (bottom two datasets) correspond to decaying pulses, determined by the position of the meeting point relative to the steady-state point (black square), as predicted by Eq.~\eqref{eq:growing_decaying}. (inset) The dimensionless growth rate, $\dot{L}(t)/c_{\rm r}(t)$, of the two growing pulses (same symbols and colors as in the main panel) vs.~$(t-t_0)/{\cal T}$ for $t\!\ge\!t_0$, with ${\cal T}\!=\!L^{(\rm ss)}\!/c_{\rm r}^{(\rm ss)}$ corresponding to the fixed $\tau$. The results confirm Eq.~\eqref{eq:slowness}, demonstrating the slow evolution of growing pulses, i.e., their sustained nature.}
    \label{fig:fig1}
\end{figure*}

\vspace{-0.5cm}
\section{T\lowercase{he two-dimensional theory of slip pulses}}

Consider the frictional system illustrated in Fig.~\ref{fig:fig1}a, where an $x\!-\!y$ coordinate system and the ambient stress conditions (normal prestress $\sigma$ and shear prestress $\tau$) are defined. The release of the energy stored in the blocks is initiated at a point along the fault, i.e., at the hypocenter (Fig.~\ref{fig:fig1}a). This triggering (nucleation) event gives rise to sliding motion, mediated by a propagating rupture that is accompanied by a complex in-plane displacement field ${\bm u}(x,y,t)$. The latter emerges as a solution of~\cite{das1980numerical}
\begin{equation}
\label{eq:BIM}
    \sigma\,f[v(x,t),\,\ldots] = \tau -\frac{\mu}{2c_{\rm s}}v(x,t) + s(x,t) \ ,
\end{equation}
where the left-hand-side (LHS) is the frictional strength that balances the interfacial shear stress in the right-hand-side (RHS). The frictional strength is proportional to $\sigma$ and depends on the slip rate $v(x,t)$ through the functional $f[v(x,t),\,\ldots]$, the fault constitutive law (``friction law''). The latter also depends on additional fields, schematically represented by the ellipsis.

The slip rate $v(x,t)\!\equiv\!\dot{\delta}(x,t)$ is the partial time derivative (superposed dot) of the slip $\delta(x,t)\!\equiv\!u_x(x,y\=0^+,t)-u_x(x,y\=0^-,t)$, i.e., the displacement discontinuity across the fault. The dependence of the frictional strength on $v$ is a generic property of frictional interfaces~\cite{Tullis1986,Kilgore1993,Marone1998a,Nakatani2001,Baumberger2006Solid}. In addition, the frictional strength depends on the state of the fault, which is generally slip history dependent, and possibly on other fields. The state of the fault may account for the real contact area between the two block, which is generally much smaller than the nominal contact area due to small-scale roughness~\cite{Dieterich1994a,Baumberger2006Solid,Rubinstein2004} (as in the rate-and-state friction framework~\cite{Dieterich1979,Ruina1983,Marone1998a,Nakatani2001,Baumberger2006Solid}) or, e.g., for the presence of a gouge layer~\cite{Marone1990}. Other fields may include fluid pore pressure and temperature (defined also off-fault, i.e., in the bulks/blocks), as in the thermal pressurization model, a widely used constitutive law in earthquake modeling~\cite{Rice2006,Viesca2015,brantut2019stability}. Each of these fields satisfies its own evolution equation, which is coupled to Eq.~\eqref{eq:BIM}. We do not specify any of these fields here (hence the ellipsis), just highlight that in a minimal description of frictional systems at least one such field has to be considered~\cite{Perrin1995,pomyalov2023self}.

The RHS of Eq.~\eqref{eq:BIM} consists of three contributions; the prestress $\tau$, a term proportional to $v(x,t)$ ($\mu$ and $c_{\rm s}$ are shear modulus and wave-speed, respectively) that accounts for wave radiation away from the fault, and $s(x,t)$, which accounts for the long-range elastodynamic interactions between different parts of the fault. It is valid when the blocks are very large and/or over times too short for reflected waves to reach the fault, as assumed here. Solutions to Eq.~\eqref{eq:BIM}, coupled to equations for additional fields represented by the ellipsis, correspond to slip pulses if $v(x,t)$ is finite over a finite length $L(t)$, i.e., $v(x,t)\!\to\!0$ both ahead and behind the rupture. An example of a pulse is shown in Fig.~\ref{fig:fig1}b, featuring an instantaneous propagation speed $c_{\rm r}(t)$, size $L(t)$ and peak slip rate $v_{\rm max}(t)$. The theoretical challenge is to predict the spatiotemporal evolution of pulses in terms of such a small set of discrete observables as a function of the prestress, rupture nucleation conditions and fault properties.

In presenting the theory to be experimentally tested below, we follow the recent developments in~\cite{Brener2018,brantut2019stability,pomyalov2023self,pomyalov2023dynamics,pomyalov2024,bouchbinder2026equation}. The starting point is finding steady-state pulse solutions, i.e., solutions to Eq.~\eqref{eq:BIM} coupled to an additional field in $f[v(x,t),\,\ldots]$, in terms of a steadily co-moving coordinate $x-c_{\rm r}^{(\rm ss)}t$, for a fixed $\sigma$. Finding a family of such steady-state pulse solutions for a fixed $\sigma$ and various $\tau$ values results in $L^{(\rm ss)}(\tau)$, $v_{\mbox{\tiny \rm max}}^{(\rm ss)}(\tau)$ and $c_{\rm r}^{(\rm ss)}(\tau)$. In Fig.~\ref{fig:fig1}c, we show an example of such a family of steady-state anti-plane solutions (dashed line), obtained in the framework of rate-and-state friction~\cite{pomyalov2023self,pomyalov2023dynamics} and presented in the $L-v_{\rm{max}}$ plane. That is, we plot $v_{\mbox{\tiny \rm max}}^{(\rm ss)}(\tau)$ against $L^{(\rm ss)}(\tau)$, parameterized by $\tau$, which gives rise to a $v_{\mbox{\tiny \rm max}}^{(\rm ss)}[L^{(\rm ss)}]$ line.
\begin{figure*}[ht]
\centering
\includegraphics[width=1.9\columnwidth]{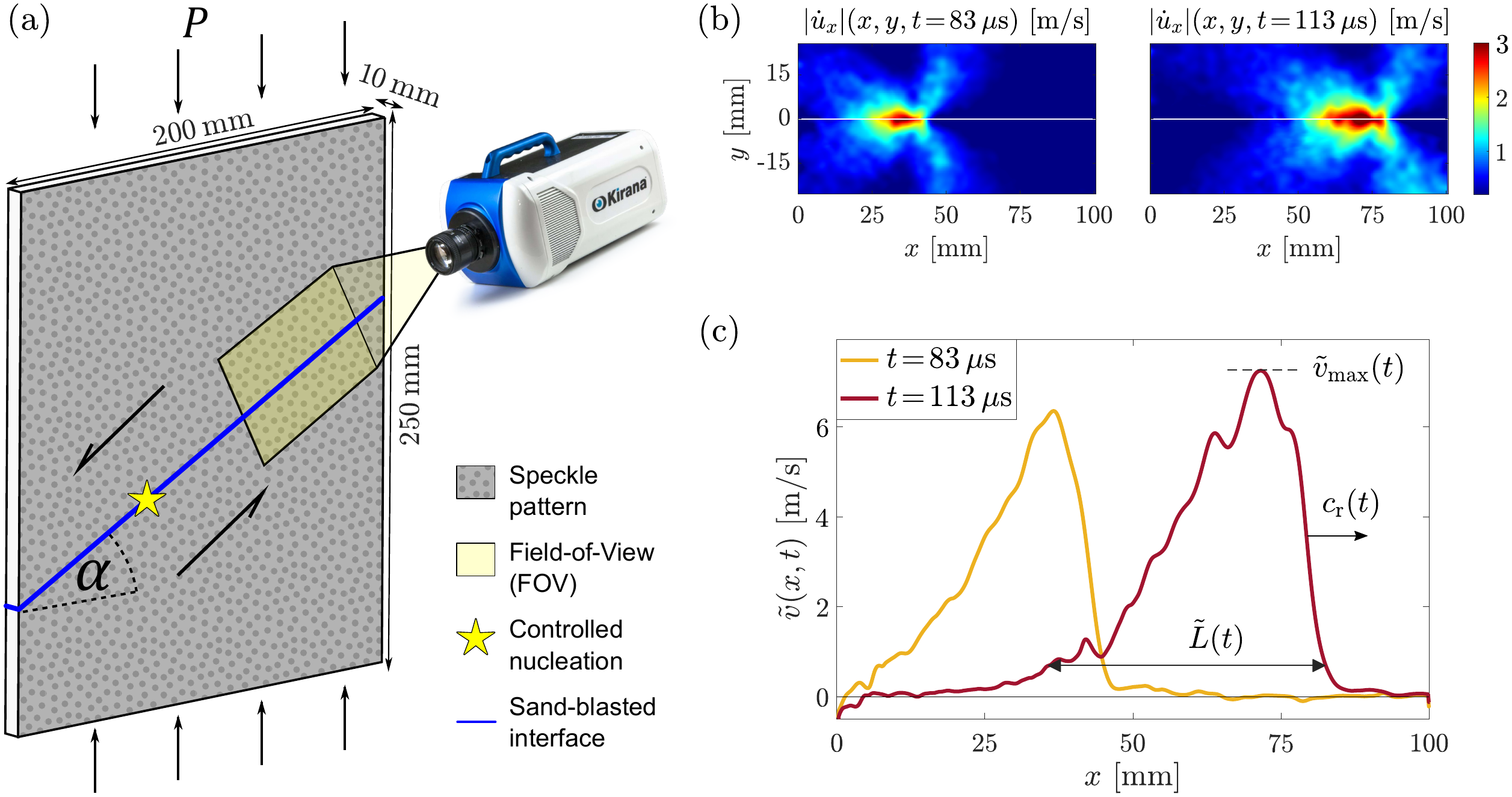}
\caption{\footnotesize{\bf Experimental setup and measured slip pulses.} (a) A sketch of the experimental system mimicking a natural fault. It is composed of two, quasi-2D (physical dimensions are stated) PMMA plates, forming a nominally flat interface/fault (blue plane) at an angle $\alpha$ relative to a direction perpendicular to an applied compressive stress of magnitude $P$. These two parameters allow to carefully control the prestress conditions on the fault through the relations $\tau\!=\!\tfrac{1}{2}P \sin(2\alpha)$ and $\sigma\!=\!P \cos^2(\alpha)$. The nominally flat fault features small-scale roughness that is controlled by a particle blasting procedure (`sand-blasting'). Rupture is nucleated/triggered in a localized region (yellow star) on the fault using a voltage-controlled burst that leads to a local release of energy (Supplementary Materials). The spatiotemporal properties of the nucleated rupture are probed using a ultrahigh-speed camera, whose Field-of-View (FOV) is located somewhat away from the nucleation site to allow rupture to become sufficiently well developed when entering the FOV. The in-plane displacement field ${\bm u}(x,y,t)$ is measured using digital image correlation, enabled by a printed speckle pattern (Supplementary Materials). The fault is located at $y\!=\!0$, where $y$ is the fault-normal coordinate and $x$ is the fault-parallel coordinate, as in Fig.~\ref{fig:fig1}a. (b) An example of $|\dot{u}_x|(x,y,t)$ of a growing pulse, propagating along the positive $x$ axis at two times $t$ after nucleation. (c) The experimental slip rate field $\tilde{v}(x,t)$ obtained from the measured displacement field (Supplementary Materials) at the same time snapshots presented in (a). $\tilde{v}(x,t)$ is used to extract the basic pulse quantities $c_{\rm r}(t)$ (propagation speed, in the sub-Rayleigh regime), $\tilde{L}(t)$ (size) and $\tilde{v}_{\rm max}(t)$ (peak slip rate), see text for discussion and additional details.}
    \label{fig:fig2}
\end{figure*}

It has been then shown that each point along the line is unstable, i.e., that steady-state slip pulses are categorically unstable, independently of the value of the prestress $\tau$. Yet, and quite remarkably, it has been shown that while the steady-state $v_{\mbox{\tiny \rm max}}^{(\rm ss)}[L^{(\rm ss)}]$ line consists of unstable points for different $\tau$ values, it serves as a dynamic attractor for the evolution of unsteady pulses under fixed prestresses $\tau$ and $\sigma$. That is, it is predicted that unsteady pulses evolve such that their time-dependent peak slip rate $v_{\rm max}(t)$ and size $L(t)$ are related according to~\cite{pomyalov2023dynamics,bouchbinder2026equation}
\begin{equation}
\label{eq:eom}
    v_{\rm max}(t) = v_{\mbox{\tiny \rm max}}^{(\rm ss)}[L^{(\rm ss)}\=L(t)] \ ,
\end{equation}
which can be viewed as an implicit equation of motion of slip pulses. In fact, it has been shown that the structure of Eq.~\eqref{eq:eom} is valid for any pair of discrete pulse observables/properties, e.g., $c_{\rm r}(t)\=c_{\rm r}^{(\rm ss)}[L^{(\rm ss)}\=L(t)]$, if the time-dependent pulse propagation speed $c_{\rm r}(t)$ and $L(t)$ are considered.

The above predictions imply that the dynamics of unsteady pulses are described by a single time-dependent function, from which all other pulse properties are obtained using a family of steady-state pulse solutions. Moreover, it has been predicted that unsteady pulses that reside above their corresponding $\tau$ steady-state point are growing pulses, $\dot{L}(t)\!>\!0$, and those that reside below it are decaying pulses, $\dot{L}(t)\!<\!0$, i.e.,
\begin{equation}
\label{eq:growing_decaying}
\dot{L}(t)\,\,
\begin{cases}
  > 0 \qquad\hbox{for}\quad L(t)\!>\!L^{(\rm ss)}(\tau) \ , \\[2mm]
  < 0 \qquad\hbox{for}\quad L(t)\!<\!L^{(\rm ss)}(\tau) \ .
\end{cases}
\end{equation}
That is, the steady-state point $(L^{(\rm ss)},v_{\mbox{\tiny \rm max}}^{(\rm ss)})$ at a given prestress $\tau$ (and $\sigma$) splits the steady-state $v_{\mbox{\tiny \rm max}}^{(\rm ss)}[L^{(\rm ss)}]$ line into growing and decaying pulses~\cite{pomyalov2023dynamics,bouchbinder2026equation}.

Equations~\eqref{eq:eom}-\eqref{eq:growing_decaying} (and their counterparts for other pairs of pulse properties) have been numerically verified~\cite{pomyalov2023dynamics}, see the rate-and-state friction example in Fig.~\ref{fig:fig1}c. These predictions are supplemented by yet another prediction regrading growing pulses, suggesting that they grow slowly in the sense that~\cite{pomyalov2023dynamics,bouchbinder2026equation}
\begin{equation}
\label{eq:slowness}
 0 < \dot{L}(t)/c_{\rm r}(t) \ll 1 \ .
\end{equation}
The latter, which is confirmed numerically in the inset of Fig.~\ref{fig:fig1}c, implies that growing pulses propagate several times their characteristic size without appreciably changing their properties, i.e., they are in fact `sustained pulses'~\cite{pomyalov2023dynamics,bouchbinder2026equation}.

In additional to the above-mentioned rate-and-state friction numerical confirmation of Eqs.~\eqref{eq:eom}-\eqref{eq:slowness}, obtained under anti-plane shear conditions~\cite{pomyalov2023dynamics}, an explicit analytic counterpart of Eq.~\eqref{eq:eom} has been obtained~\cite{bouchbinder2026equation} and verified in this case. There are also strong indications that these relations are satisfied for pulses driven by thermal pressurization of pore fluids~\cite{brantut2019stability}, which is a very different constitutive law. Therefore, we expect Eqs.~\eqref{eq:eom}-\eqref{eq:slowness} to be valid under in-plane conditions for a broad range of realistic fault constitutive laws.\\

\vspace{-0.5cm}
\section{E\lowercase{xperimental validation}}

We present here direct experimental support to the above-discussed theory of slip pulses. The experimental setup (Fig.~\ref{fig:fig2}a and Supplementary Materials) is similar to the one recently employed in~\cite{Rubino2017,rubino2019full,tal2020illuminating,rubino2020spatiotemporal,gabrieli2025lab} to mimic natural earthquakes in the lab. It involves two quasi-2D poly(methylmethacrylate) (PMMA) plates, forming a dry, gouge-free, nominally flat interface (fault) once externally pressed together. The loading configuration allows to carefully control the fault prestress levels $\sigma$ and $\tau$ (Fig.~\ref{fig:fig2}a).

The fault features small-scale roughness, produced by a particle blasting procedure (Supplementary Materials), which affects the effective fault constitutive law. Rupture is nucleated/triggered in a localized region on the fault using a voltage-controlled burst that leads to a local release of energy (Fig.~\ref{fig:fig2}a and Supplementary Materials). By varying the voltage $V$, the intensity of rupture nucleation is controlled. Overall, our experimental approach enables to control the prestress level and rupture nucleation conditions, as well as to vary the fault constitutive law (through the amplitude of small-scale roughness), all directly relevant for testing the theory.

The spatiotemporal properties of the nucleated rupture are probed by combining ultrahigh-speed photography and digital image correlation (Fig.~\ref{fig:fig2}a and Supplementary Materials). An example of $|\dot{u}_x|(x,y,t)$ of a propagating pulse, at two times after nucleation, is presented in Fig.~\ref{fig:fig2}b. This pulse, like all other experimental pulses discussed in this work, features $c_{\rm r}(t)$ in the sub-Rayleigh regime. The corresponding slip rate field $\tilde{v}(x,t)$, at the same two times, is presented in Fig.~\ref{fig:fig2}c. The superposed tilde represents the experimentally measured field (Supplementary Materials) and we denote the peak slip rate and pulse size extracted from $\tilde{v}(x,t)$ by $\tilde{L}(t)$ and $\tilde{v}_{\rm max}(t)$, respectively (Fig.~\ref{fig:fig2}c). With these at hand, we are ready to test the theory.

We start by testing Eqs.~\eqref{eq:eom}-\eqref{eq:growing_decaying}. Note that the exact fault constitutive law for the experimental system is not known, hence one cannot a priori compute the exact fixed-$\sigma$ steady-state $\tilde{v}_{\mbox{\tiny \rm max}}^{(\rm ss)}[\tilde{L}^{(\rm ss)}]$ line. Yet, Eq.~\eqref{eq:eom} predicts that the dynamics of slip pulses triggered under fixed $\sigma$ and small-scale interfacial roughness, specifically their time-dependent $\tilde{L}(t)$ and $\tilde{v}_{\rm{max}}(t)$, will form a monotonically increasing line (a dynamic attractor) when presented in the $\tilde{L}-\tilde{v}_{\rm{max}}$ plane, similarly to the theoretical dashed line in Fig.~\ref{fig:fig1}c. Moreover, Eq.~\eqref{eq:growing_decaying} predicts that pulses along this line split into growing (hot-colored circles in Fig.~\ref{fig:fig1}c) and decaying (cold-colored diamonds therein) pulses, depending on their position relative to the steady-state point at a fixed $\tau$ (black square therein).
\begin{figure*}[ht!]
    \centering
    \includegraphics[width=2.10\columnwidth]{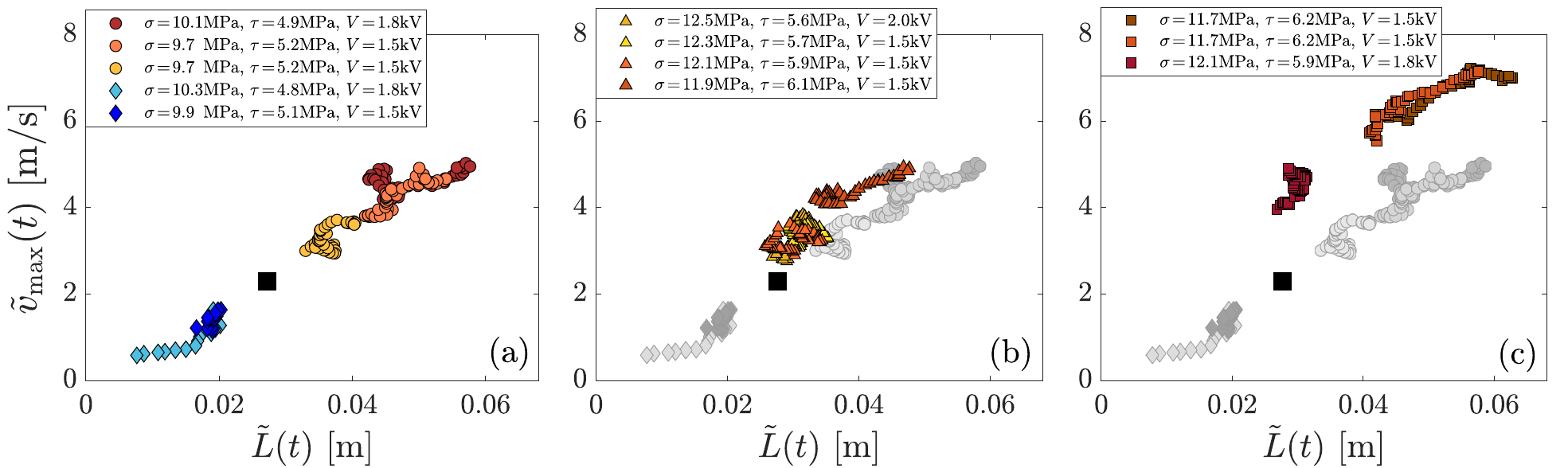}
    \caption{\footnotesize{\bf Experimental validation of the theory.} (a) Experimental results for $\tilde{v}_{\rm{max}}(t)$ vs.~$\tilde{L}(t)$ of $5$ slip pulses obtained with $\sigma\!=\!10\pm 0.3$ MPa, $\tau\!=\!5 \mp 0.2$ MPa and two values of the nucleation voltage $V$ (see legend). Growing pulses (circles) appear in hot colors and decaying ones (diamonds) appear in cold colors, following the symbol and color codes of the theoretical Fig.~\ref{fig:fig1}c. The black square is an estimate of the location of the relevant steady-state point, separating decaying and growing pulses. See text for an extensive discussion of the experimental results and their relations to the theoretical predictions. (b) The same as (b), but for growing pulses obtained with $\sigma\!=\!12.2\pm 0.3$ MPa, $\tau\!=\!5.85 \mp 0.25$ MPa and two values of the nucleation voltage $V$ (see legend), see text for extended discussion. The results of (a) are superposed with light grayscale colors for comparison. The experiments in (a) and (b) have been performed with the same small-scale fault roughness. (c) The same as (a), but for growing pulses obtained with $\sigma\!=\!11.9 \pm 0.2$ MPa, $\tau\!=\!6.05 \mp 0.15$ MPa and two values of the nucleation voltage $V$ (see legend), and a reduced amplitude of the small-scale fault roughness. See text for extended discussion.}
    \label{fig:fig3}
\end{figure*}

In Fig.~\ref{fig:fig3}a, we first superpose the $\tilde{v}_{\rm{max}}(t)$ vs.~$\tilde{L}(t)$ results of $3$ experiments giving rise to growing pulses (circles, hot colors), performed under approximately fixed $\sigma$, with a comparably small variation in $\tau$ and two values of the nucleation voltage $V$ (see legend). The $3$ growing pulses approximately form a monotonic line in the $\tilde{L}-\tilde{v}_{\rm{max}}$ plane, as predicted. We then superpose the results of $2$ additional experiments giving rise to decaying pulses (diamonds, cold colors), obtained under approximately the same $\sigma$ and slightly reduced $\tau$ compared to two of the growing pulses (see legend).

The $2$ decaying pulses not only approximately form a monotonic line in the $\tilde{L}-\tilde{v}_{\rm{max}}$ plane, but this line is also a continuation of the reduced-dimensionality dynamical attractor followed by the $3$ growing pulses. These results also indicate that the steady-state point corresponding to the average $\langle\sigma\rangle$ and $\langle\tau\rangle$ values resides in between these two subsets (see superposed black square). The experimental results in Fig.~\ref{fig:fig3}a closely mirror the theoretical results in Fig.~\ref{fig:fig1}c and as such provide significant support to the predictions in Eqs.~\eqref{eq:eom}-\eqref{eq:growing_decaying}.

To test the generality of these findings, we focus on growing pulses --- that are of great dynamical importance --- and superpose in Fig.~\ref{fig:fig3}b the $\tilde{v}_{\rm{max}}(t)$ vs.~$\tilde{L}(t)$ results of $4$ experiments (triangles, hot colors), performed under somewhat higher $\langle\sigma\rangle$ and $\langle\tau\rangle$ values (see legend) compared to those in Fig.~\ref{fig:fig3}a (still within the range of existence of slip pulses). The theoretical expectation is that these growing pulses will follow a slightly shifted reduced-dimensionality dynamical attractor. Indeed, it is observed that the $4$ growing pulses approximately form a monotonic line in the $\tilde{L}-\tilde{v}_{\rm{max}}$ plane, which is slightly shifted upwards compared to the results presented in Fig.~\ref{fig:fig3}a (those are superposed with reduced size symbols and gray colors), yet again supporting the theoretical predictions.

The results in Figs.~\ref{fig:fig3}a-b were obtained for faults featuring the same small-scale roughness (Supplementary Materials). Next, we test the generality of the theoretical predictions with respect to the fault constitutive law, achieved by varying the amplitude of the small-scale roughness. Specifically, we reduced the latter (Supplementary Materials), resulting in a smoother fault, which is expected to give rise to a reduced frictional resistance. In turn, larger slip rates should emerge, leading to an upward shift in the dynamical attractor in the $\tilde{L}-\tilde{v}_{\rm{max}}$ plane. In Fig.~\ref{fig:fig3}c, we superpose the results of $3$ experiments (squares, hot colors) performed with smoother faults and approximately fixed $\sigma$ and $\tau$ (see legend). It is observed that the $3$ growing pulses form a monotonic line that is indeed shifted upwards (compared to the results for rougher faults, superposed for comparison), further supporting the theoretical predictions.

Next, we test the prediction in Eq.~\eqref{eq:slowness}, which goes beyond the predictions tested in Fig.~\ref{fig:fig3} as it imposes a constraint on the time evolution of growing pulses as they move along the dynamical attractor in the $\tilde{L}-\tilde{v}_{\rm{max}}$ plane (or any other pair of pulse properties). Specifically, while the precise time evolution of growing pulses may depend to some extent on the specific fault constitutive law~\cite{bouchbinder2026equation}, Eq.~\eqref{eq:slowness} predicts that it is generally slow in the sense that the dimensionless ratio $\dot{L}(t)/c_{\rm r}(t)$ is small at all times. This prediction is of great dynamical importance as it suggests that growing pulses propagate several times their characteristic size without appreciably changing their properties, i.e., they are `sustained pulses'~\cite{Brener2018,pomyalov2023dynamics}, which explains why pulses retain their compact nature in fault dynamics.

\begin{figure}[ht!]
    \centering
    \includegraphics[width=0.95\columnwidth]{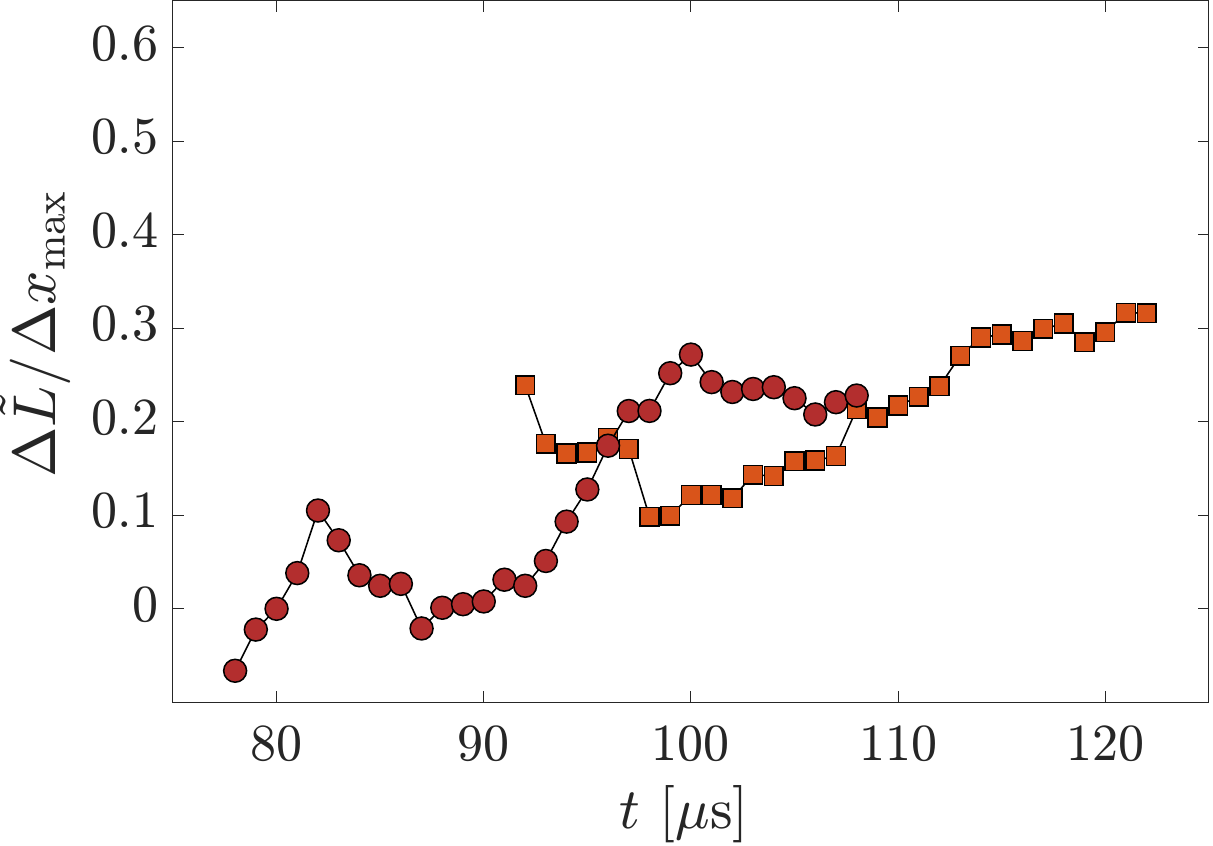}
    \caption{\footnotesize\textbf{The slow evolution of growing pulses.} $\Delta{\tilde L}(t)/\Delta{x}_{\rm max}(t)$ of two growing pulses presented in Fig.~\ref{fig:fig3} (one from panel (a) and the other from panel (c) therein, using the same symbols and colors). The experimental results support the theoretical prediction in Eq.~\eqref{eq:slowness} through its integral corollary $\Delta{\tilde L}(t)/\Delta{x}_{\rm max}(t)\!\ll\!1$ (see text and Supplementary Materials), demonstrating the slow evolution of growing pulses (featured by all observed growing pulses). See text for additional discussion.}
    \label{fig:fig4}
\end{figure}

Directly testing Eq.~\eqref{eq:slowness} with the experimental data (using $\tilde{L}(t)$ as a proxy for $L(t)$) involves taking time derivatives of measured quantities, i.e., $\dot{\tilde{L}}(t)$ and $c_{\rm r}(t)\=\dot{x}_{\rm max}(t)$, where $x_{\rm max}(t)$ is the interfacial location of the slip rate peak, $\tilde{v}_{\rm{max}}(t)$. Yet, measurement noise (Supplementary Materials) renders the estimation of experimental time derivatives noisy. Alternatively, one can obtain an integral corollary of Eq.~\eqref{eq:slowness}, which takes the form $\Delta{\tilde L}(t)/\Delta{x}_{\rm max}(t)\!\ll\!1$ (Supplementary Materials). Here, $\Delta{\tilde L}(t)\={\tilde L}(t)\!-\!{\tilde L}(t_0)$ and $\Delta{x}_{\rm max}(t)\=x_{\rm{max}}(t)-x_{\rm{max}}(t_0)$, where $t_0$ is the first time point of each growing pulse presented in Fig.~\ref{fig:fig3}. This prediction is confirmed in Fig.~\ref{fig:fig4} for two growing pulses presented in Fig.~\ref{fig:fig3} for two different faults (see legend), directly demonstrating the slow evolution of growing pulses.

\vspace{-0.5cm}
\section{D\lowercase{iscussion}}

Our experimental results confirm essentially all the predictions of the recently developed theory of slip pulses, also demonstrating its generality across prestress conditions and fault constitutive laws. The theory provides a solid, quantitative ground for understanding and analyzing slip pulses in diverse frictional systems and phenomena, ranging from stick-slip instabilities in engineering structures~\cite{Persson1998} to the analysis of crustal earthquakes~\cite{Scholz2002}. In the latter context, the theory may pave the way for an improved estimation of source parameters of pulse-like earthquakes and the interpretation of observational data. Moreover, since slip pulses are believed to significantly contribute to the emergence of fault slip complexity due to their unstable and compact nature~\cite{brantut2019stability,roch2024finite,pomyalov2024}, the established theory is expected to lead to developments in this direction as well.

Finally, note that we focused in this work on slip pulses, mostly growing but also decaying, which constitute two dynamic rupture styles. A decaying-to-growing pulse transition has been discussed in Fig.~\ref{fig:fig3}a and is explicitly demonstrated in Additional Fig.~1 below. In addition to these rupture styles, as explained in the introduction, crack-like rupture that features a slipping zone that increases with the rupture propagation distance can also be triggered, depending on the prestress and nucleation conditions~\cite{Gabriel2012,Brener2018}. In fact, the very same theoretical framework discussed here also predicts a pulse-to-crack transition~\cite{Brener2018}, e.g., by increasing the nucleation intensity at fixed prestress conditions. An example for such a transition is provided in Additional Fig.~2 below.

\vspace{1.3cm}
\hspace{-0.4cm}{\bf \large Acknowledgements}

E.B.~acknowledges support from the Minerva Foundation (with funding from the Federal German Ministry for Education and
Research) and the Harold Perlman family. T.G.~and Y.T.~acknowledge support from the Israel Science Foundation (ISF Grant No.~2188/24). A.S.~and E.B.~thank Anna Pomyalov and Fabian Barras for helpful discussions.

\clearpage

\onecolumngrid
\begin{center}
	\textbf{\large Additional Figures}
\end{center}
\renewcommand{\figurename}{Additional Fig.}
\setcounter{figure}{0}
\begin{figure}[ht!]
   \includegraphics[width = 0.75\textwidth]{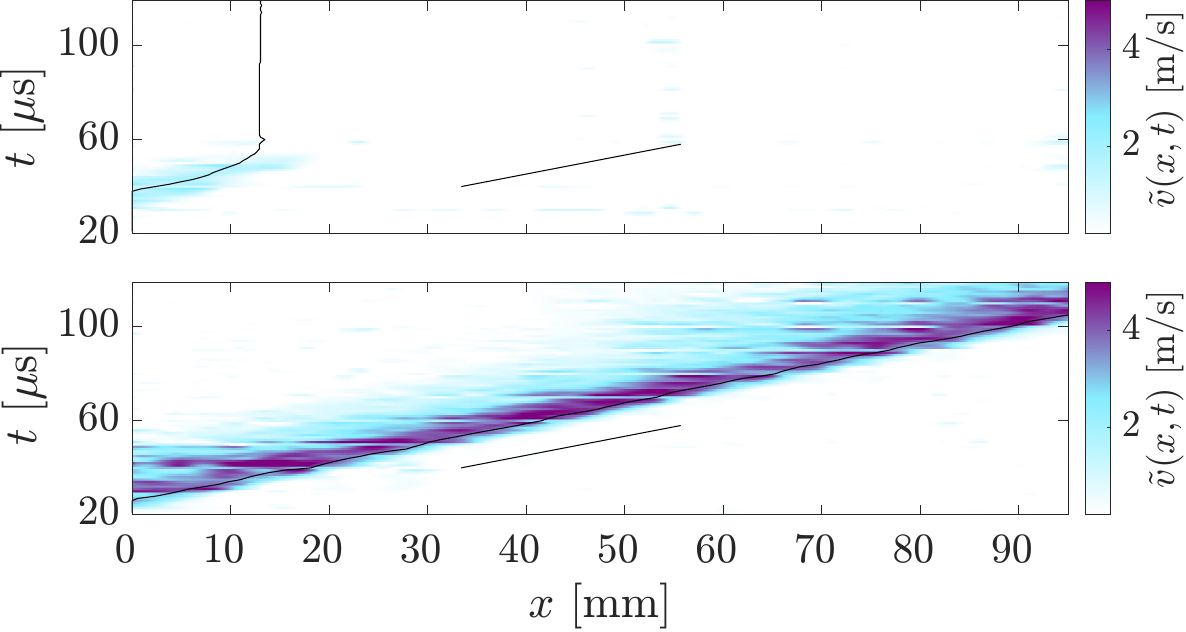}
  \caption{\footnotesize \textbf{An example of a decaying-to-growing pulse transition.} (top) A space-time plot of $\tilde{v}(x,t)$ of a decaying pulse shown in Fig.~\ref{fig:fig3}a (obtained with $\sigma\!=\!10.3$ MPa, $\tau\!=\!4.8$ MPa and $V\!=\!1.8$ kV, see legend therein). The thin superposed line tracks the position of the pulse, highlighting the decay and eventually the arrest of the pulse. The tilted straight line (the same as in the bottom panel) corresponds to the Rayleigh wave-speed $c_{_{\rm R}}\!=\!1237$ m/s. (bottom) The same as the top panel, but for a growing pulse (also shown in Fig.~\ref{fig:fig3}a, obtained with $\sigma\!=\!10.1$ MPa, $\tau\!=\!4.9$ MPa and $V\!=\!1.8$ kV, see legend therein). This decaying-to-growing pulse transition is induced by a small change of the prestress at a fixed nucleation intensity (nucleation voltage $V$). The growing pulse propagation speed $c_{\rm r}(t)\!<\!c_{_{\rm R}}$ can be obtained from the slope of the thin superposed line, demonstrating its sub-Rayleigh nature (compare to the slope of the tilted straight line, which corresponds to $c_{_{\rm R}}$), like all ruptures considered in this work.}
  \label{fig:Ext_Data_fig1}
\end{figure}
\begin{figure}[ht!]
   \includegraphics[width = 0.75\textwidth]{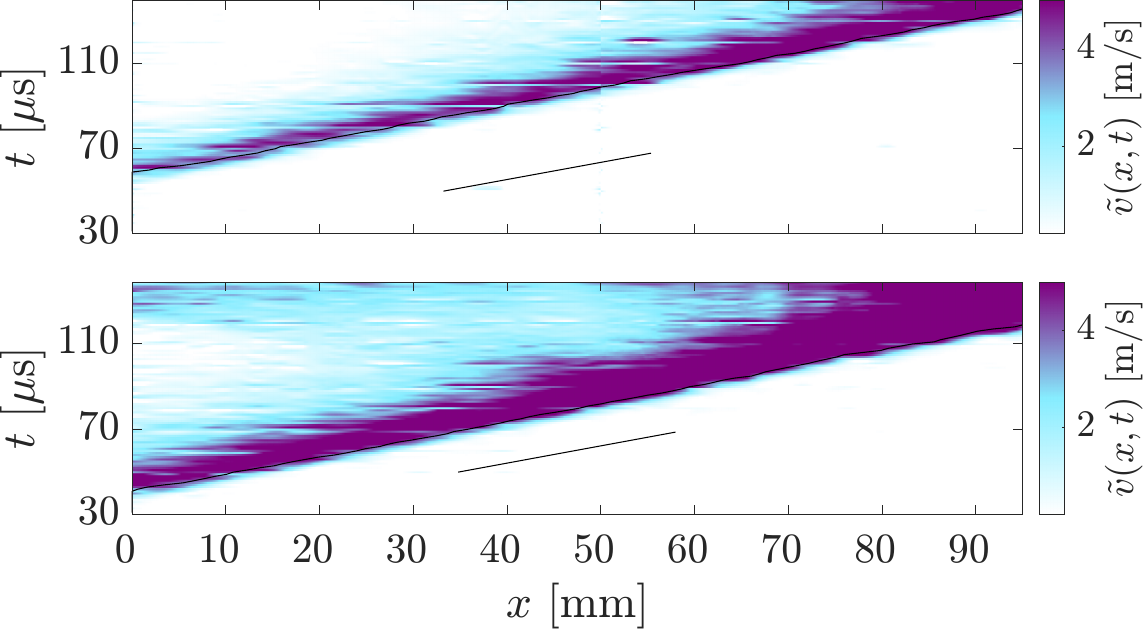}
  \caption{\footnotesize \textbf{An example of a pulse-to-crack transition.} (top) The same as Fig.~\ref{fig:Ext_Data_fig1}(bottom), but for a growing pulse obtained with $\sigma\!=\!11.5$ MPa, $\tau\!=\!6.4$ MPa and $V\!=\!1.3$ kV. (bottom) The same as the top panel, but for a propagating crack-like rupture obtained with the same prestress and a more intense nucleation associated with $V\!=\!1.5$ kV. Note the nearly constant slip rate of $\sim\!2$ m/s left behind the rupture edge at late time, the characteristic/defining property of crack-like rupture (to be contrasted with the vanishing slip behind a pulse-like rupture, see top panel. This pulse-to-crack transition is induced by increasing the nucleation intensity (nucleation voltage $V$) at a fixed prestress.}
  \label{fig:Ext_Data_fig2}
\end{figure}

\clearpage

\onecolumngrid
\begin{center}
	\textbf{\large Supplementary Materials}
\end{center}
\renewcommand{\figurename}{Fig.}

\setcounter{equation}{0}
\setcounter{figure}{0}
\setcounter{section}{0}
\setcounter{table}{0}
\makeatletter
\renewcommand{\theequation}{S\arabic{equation}}
\renewcommand{\thefigure}{S\arabic{figure}}
\renewcommand{\thesection}{S-\arabic{section}}
\renewcommand{\thetable}{S-\arabic{table}}

\twocolumngrid

The goal of this document is to provide additional technical details about the results presented in the manuscript.

\vspace{-0.5cm}
\section{E\lowercase{xperimental system}}
\label{sec:exp_system}

Here we provide information about the experimental system~\cite{Rubino2017,rubino2019full,tal2020illuminating,rubino2020spatiotemporal,gabrieli2025lab}, while the analysis of the experimental data is discussed in Sect.~\ref{sec:DIC}. The experimental system, illustrated in Fig.~\ref{fig:fig2}a, is composed of two precut, quasi-2D poly(methylmethacrylate)
(PMMA) plates (see dimensions therein), forming an interface (fault) at an angle $\alpha$ relative to a direction perpendicular to the applied compressive stress of magnitude $P$. These two parameters allow to carefully control the prestress conditions on the fault through the relations $\tau\=\tfrac{1}{2}P \sin(2\alpha)$ and $\sigma\=P \cos^2(\alpha)$. PMMA is used as an analog material since it exhibits a low shear modulus compared to crustal rocks, which enables us to explore well-developed ruptures within samples of tens of cm, significantly smaller than those that would have been required for rocks.

The fault is nominally flat, yet featuring small-scale roughness that is varied through a post-cutting interfacial preparation protocol, involving a particle blasting procedure (also known as abrasive blasting or `sandblasting'). Specifically, aluminum oxide particles of either a mean diameter of $180\,\mu$m (Fig.~\ref{fig:fig3}a-b) or approximately in the range of $125\!-\!150\,\mu$m (Fig.~\ref{fig:fig3}c) are used, where smaller particles give rise to a reduced small-scale roughness (smoother fault). Rupture is nucleated in a localized region on the fault by triggering a burst of a NiCr wire, whose intensity/amplitude is controlled by the voltage $V$ applied to it. The experimental control parameters $\alpha$, $P$ and $V$ are selected such that under most circumstances pulse-like rupture is triggered (but see the crack-like rupture discussed in Additional Fig.~2(bottom)).

The spatiotemporal dynamics of the triggered rupture are captured using a dynamic imaging technique that combines ultrahigh-speed
photography (Specialized Imaging, Kirana 5M camera~\cite{Crooks2013}) with a frame rate of $10^6$ fps and Digital Image Correlation~\cite{Sutton2009} (DIC, Vic-2D software by Correlated Solutions). DIC is enabled by spraying white matte paint on the
face of the PMMA plates and printing a digitally produced pattern of black dots with a Yotta YD-F2513R5 Digital UV Flatbed Printer. The dots are randomly distributed with similar density all over the face, and their number is such that $50\%$ of the face is black due to
the dots and the rest is a white matte background, generating a suitable speckle pattern.

The camera features a spatial resolution of $924\,(W) \times 768\,(H)$ pixel$^2$, and it was positioned to provide a $10 \times 8$ cm$^2$ Field-of-View (FOV), suitable for capturing the evolution of pulses of typical width of about $\tilde{L}\!\simeq\!5$ cm. The FOV was positioned within $1.5\!-\!4$ cm from the nucleation site, where the upper end of the range was typically more suitable for capturing growing pulses (to allow them to form/mature after nucleation and before observation) and the lower end was more suitable for capturing decaying pulses. The in-plane displacement field ${\bm u}(x,y,t)\=[u_x(x,y,t), u_y(x,y,t)]$ is obtained by DIC employing the aforementioned speckle pattern. DIC and noise reduction procedures are extensively discussed next.

\vspace{-0.5cm}
\section{D\lowercase{igital \uppercase{I}mage \uppercase{C}orrelation and the subset size}}
\label{sec:DIC}

\begin{figure*}[thb]
\includegraphics[width=1.03\textwidth]{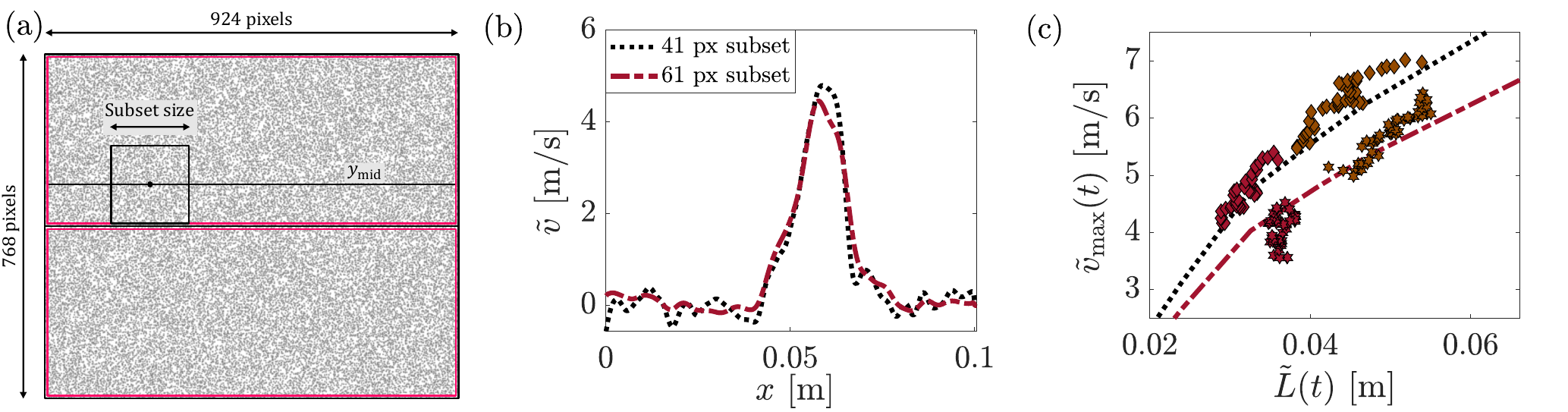}
\caption{\footnotesize (a) A sketch of an imaged frame, including the DIC analysis domains in the upper and lower halves of the frame (pink rectangles) and a subset (square) near the fault. For visual clarity, the presented subset size is $3$ times larger than its actual size. The closest point that a displacement can be assigned to using DIC is at $y_{\rm{mid}}$ (marked by a horizontal line), from which the displacement is extrapolated towards the fault, as explained in the text. The slip rate is defined approximately 2 pixels from the fault, at the upper/lower boundaries of the lower/upper analysis domains, respectively. The speckle pattern (see text) is shown at the background. (b) A slip rate field $\tilde{v}(x,t)$ snapshot for a subset size of 41 pixels (black dotted line) and of 61 pixels (brown dashed-dotted line). $\tilde{v}(x,t)$ for the 41 pixels subset features a higher $\tilde{v}_{\rm{max}}$ and a smaller $\tilde{L}$. (c) The resulting pulse evolution in the $\tilde{L}\!-\!\tilde{v}_{\rm{max}}$ plane for the 41 pixels (black dotted line added as a guide to the eye) and 61 pixels (brown dashed-dotted line added as a guide to the eye) subsets. The 61 pixels subset curve is lower than the 41 pixels one, but both form well-defined lines related by a shift.}
\label{fig:fig_S1}
\end{figure*}

As explained in Sect.~\ref{sec:exp_system}, the ultrahigh-speed Kirana camera~\cite{Crooks2013} obtains images at a frame rate of $10^6$ fps with a spatial resolution of $924\,(W) \times 768\,(H)$ pixel$^2$ in the presence of the digitally produced speckle pattern (see Fig.~\ref{fig:fig_S1}a). From the recorded images, displacements were calculated using Digital Image Correlation~\cite{Sutton2009} (DIC, Vic-2D software by Correlated Solutions), which correlates image subsets (tracking window, see the definition of the subset size in Fig.~\ref{fig:fig_S1}a) between a reference frame (prior to deformation) and the current, deformed frame. These displacements are assigned to the central pixel of each subset. The calculation yields 2D matrices of horizontal (fault-parallel, $u_x$) and vertical (fault-perpendicular, $u_y$) displacements for each frame. The slip rate field was subsequently calculated using finite differences. We conducted DIC analyses on the upper and lower plates separately to generate independent displacement fields above and below the fault.
\begin{figure}[ht]
\includegraphics[width=0.34\textwidth]{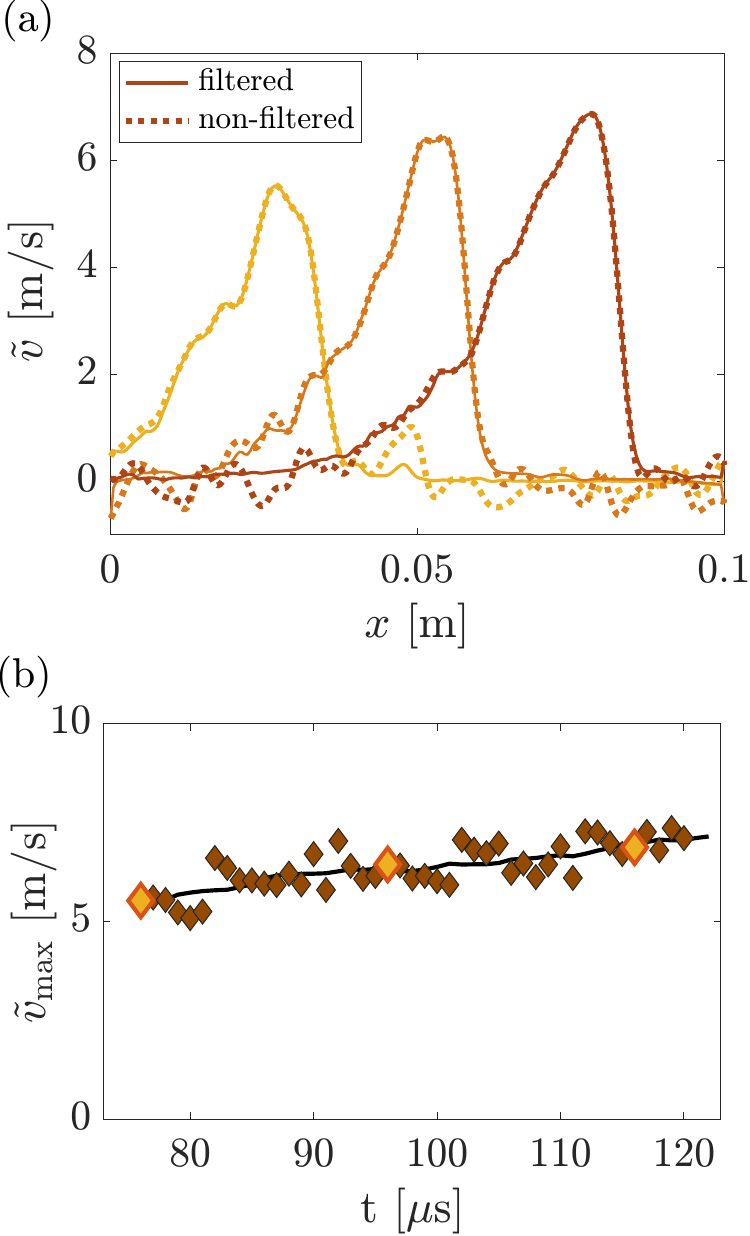}
\caption{\footnotesize (a) Pulse snapshot at $t\!=\!76$, $96$ and $116\,\mu$s (from left to right) before (dotted lines) and after (solid lines) filtering, see text for details. (b) The measured $\tilde{v}_{\rm{max}}$ (diamonds) and its moving average over $10$ frames (line). The yellow diamonds correspond to the frames presented in panel (a).}
\label{fig:fig_S2}
\end{figure}
\begin{figure}[ht]
\includegraphics[width=0.34\textwidth]{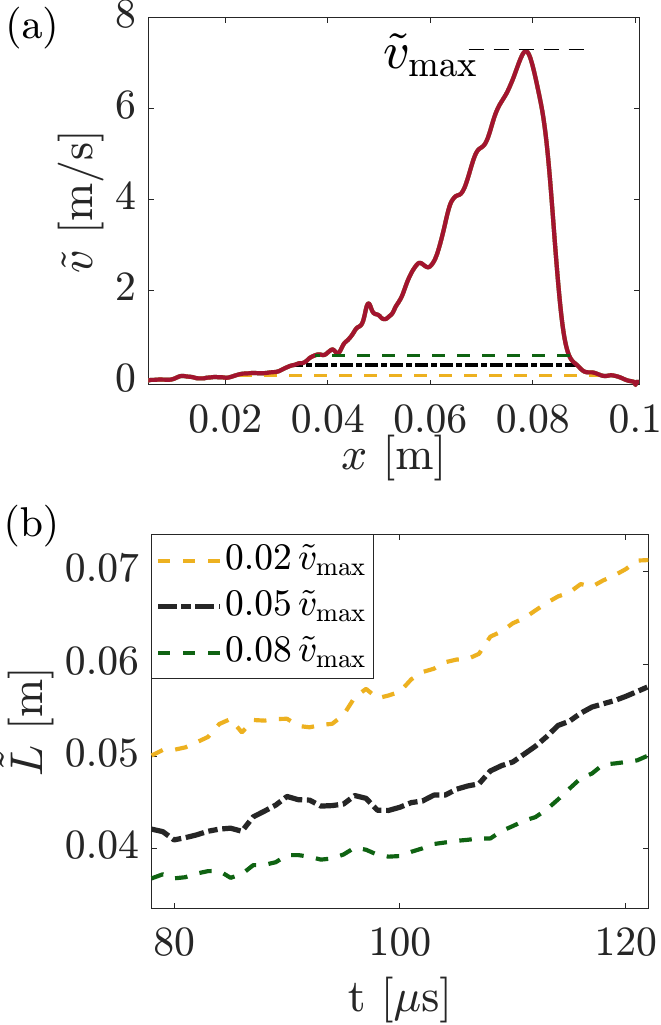}
\caption{\footnotesize (a) Pulse snapshot with three possible thresholds of $0.02\,\tilde{v}_{\rm{max}}$ (bottom), $0.05\,\tilde{v}_{\rm{max}}$ (middle) and $0.08\,\tilde{v}_{\rm{max}}$ (top) for determining $\tilde{L}$ (see also legend in panel (b)). (b) The resulting $\tilde{L}(t)$ for each threshold (see legend).}
\label{fig:fig_S3}
\end{figure}

Since the displacement of each pixel is calculated based on its surrounding subset, the analysis is possible down to half of the subset linear size, see $y_{\rm{mid}}$ in Fig.~\ref{fig:fig_S1}a. The lower edge/boundary of this limiting subset is typically located 2 pixels away from the fault. Displacements closer to these edges are extrapolated to the analysis edge using the ``Fill-Boundary'' algorithm, which is an integral feature of the Vic-2D software. The slip rate is defined as the difference in the fault-parallel particle velocity between the lower boundary of the upper subset and the upper boundary of the lower subset. Since this analysis involves extrapolation of off-fault data, it essentially involves an effective off-fault distance $y_{_0}\!>\!0$, which is much smaller than the typical pulse sizes in our experiments. That is, we have for the experimental slip rate $\tilde{v}(x,t)\!\equiv\!|v(x,y\=y_{_0},t)-v(x,y\=-y_{_0},t)|$, as used in the manuscript. Determining the actual experimental value of the effective $y_{_0}$ is of no importance, and in any case is not used per se, as long as it is small and fixed once and for all, as will be explained next.

Another implication of the displacement close to the interface being extrapolated from the displacement at the middle pixel $y_{\rm{mid}}$ is that the subset size affects the slip rate field, as demonstrated in Fig.~\ref{fig:fig_S1}b-c for two subset sizes. Clearly, the subset size also affects the effective off-fault distance $y_{_0}\!>\!0$ discussed above. A larger subset size offers a higher signal-to-noise ratio, but also involves averaging over a broader area, thereby reducing the spatial resolution and increasing the effective off-fault distance $y_{_0}$. Conversely, a smaller subset size yields a higher $\tilde{v}_{\rm{max}}$ and a smaller $\tilde{L}$ (whose extraction is discussed in Sect.~\ref{sec:noise}). The important point to note is that once the subset size is fixed (as well as the extrapolation procedure of the ``Fill-Boundary'' algorithm), pulses follow a well-defined line in the $\tilde{L}-\tilde{v}_{\rm{max}}$ plane, as demonstrated in Fig.~\ref{fig:fig_S1}c. In the manuscript, we used a fixed subset size of $41$ pixels (see Fig.~\ref{fig:fig_S1}).

\section{N\lowercase{oise reduction}}
\label{sec:noise}

Our measurement/imaging procedure features two main noise sources: a random one and a systematic, periodic one, to be discussed below. Our goal here is to discuss the noise reduction procedure applied to the measured data, leading to the experimental results reported on in the manuscript.

To mitigate the random noise, a non-local (NL) means filter was applied to the displacement fields. This filter is effective in decreasing random noise levels while preserving image structure~\cite{buades2011non,buades2008nonlocal}, and is commonly used in dynamic rupture imaging~\cite{Rubino2017,rubino2019full,gabrieli2025lab}. The NL-means algorithm replaces a pixel's value with a weighted average of the values of surrounding pixels. Higher weight is assigned to pixels with similar neighborhoods to the target pixel, as well as closer pixels. This filter is suitable for movie denoising, as well as image denoising~\cite{buades2008nonlocal}, and we applied it both spatially (in the $x\!-\!y$ domain) and temporally (in the $x\!-\!t$ domain). This procedure effectively smoothed the pulse `tails', thereby improving the accuracy of the pulse size measurements, as demonstrated in Fig.~\ref{fig:fig_S2}a. Note that the data presented in Fig.~\ref{fig:fig_S1}b-c are not filtered.

The Kirana camera also exhibits a systematic noise, which is a known feature discussed in the literature~\cite{duran2024ultrahigh,quino2021speckle,Lukic2018,Gabrieli2026}. This noise is subsequently reduced using a deniosing procedure that takes advantage of the fact that some frames suffer less from this systematic noise. Specifically, we employed a polynomial correction procedure based on the less-noisy frames. The systematic noise affects the calculated displacement field of the frames in a periodic manner (across the imaging sequence) and almost uniformly (across the frame), with the tenth frame being the noisiest, the third to fifth frames being the most accurate, and the remaining frames exhibiting varying degrees of noise levels.

In practice, we first calculated the mean displacement of every frame throughout the entire measurement sequence. We then fitted a 20th degree polynomial curve to the mean displacements of frames $3$ to $5$ in each 10-frame cycle, which sum to $54$ frames out of the total $180$ frames imaged in each experiment. We then calculated the difference between the measured mean displacements and the polynomial curve for the rest of the frames. We regard this difference as the noise, and subtract it from the displacement values of each frame. Thus, the pixels of the less-noisy frames remain approximately unchanged and the pixels of the noisy frames are changed by the difference between the mean of the frame and the value of the polynomial curve.

We repeated this denoising procedure for each half of the analysis (i.e., upper and lower plates), leading to a significant reduction in the systematic noise. However, when particle velocities are calculated using time derivatives, the residual noise is enhanced. Given the 10-frame cycle of the noise, we applied a 10-frame moving average to all extracted pulse properties. Figure~\ref{fig:fig_S2}b displays the measured maximal slip rate $\tilde{v}_{\rm{max}}$ after the polynomial correction and its corresponding moving average. The overall physical trend is of a slow growth (namely from $5$ to $7$ m/s over the given time period), as predicted theoretically (see manuscript), along with a periodic noise with a characteristic magnitude of $0.5$ m/s.

Finally, the pulse size $\tilde{L}$ (already reported above) is determined as the distance between points where the slip rate field falls below a threshold of $0.05\,\tilde{v}_{\rm{max}}$. To demonstrate the effect of this choice, we present in Fig.~\ref{fig:fig_S3}a a pulse along with three possible operational definitions of its size, corresponding to thresholds of $0.02, 0.05, 0.08\,\tilde{v}_{\rm{max}}$. The resulting sizes $\tilde{{L}}(t)$ are presented in Fig.~\ref{fig:fig_S3}b, demonstrating a similar time evolution.

\vspace{-0.5cm}
\section{Q\lowercase{uantifying the slow evolution of growing pulses}}
\label{sec:quantifying_slowness}

Growing pulses are predicted in Eq.~\eqref{eq:slowness} in the manuscript to slowly evolve in time, in the sense of $\dot{L}(t)/c_{\rm r}(t)\!\ll\!1$ (even though their propagation speed $c_{\rm r}(t)$ may be large, i.e., a sizable fraction of the Rayleigh wave-speed). In view of this property, they are termed sustained pulses, as discussed in the manuscript. Yet, directly testing the prediction $\dot{L}(t)/c_{\rm r}(t)\!\ll\!1$ involves taking time derivatives of the experimental data, which are not accurate due to measurement noise (see Sect.~\ref{sec:noise}). Therefore, in Fig.~\ref{fig:fig4} in the manuscript, we test an integral corollary of Eq.~\eqref{eq:slowness} in the form $\Delta{\tilde L}(t)/\Delta{x}_{\rm max}(t)\!\ll\!1$, where the quantities involved are defined therein (see also below). Our goal here is to briefly derive this corollary relation.

First, since $\tilde{L}(t)$ is a proxy for $L(t)$ (note that generally $\tilde{L}(t)\!\ge\!L(t)$) and since $c_{\rm r}(t)\=\dot{x}_{\rm max}(t)$ (where $x_{\rm max}(t)$ is the interfacial location of the experimental slip rate peak), Eq.~\eqref{eq:slowness} implies $\dot{\tilde{L}}(t)/\dot{x}_{\rm max}(t)\!\equiv\!\alpha(t)\!\ll\!1$. The time-dependent ratio $\alpha(t)$ is predicted to be small and bounded, i.e.~$\alpha(t)\!<\!\alpha_{\rm max}\!\ll\!1$. Next, using $\alpha_{\rm max}$, we obtain an upper bound on $\dot{\tilde{L}}(t)$ in the form $\dot{\tilde{L}}(t)\!\ll\!\alpha_{\rm max}\,\dot{x}_{\rm max}(t)$. Integrating the latter and using $\alpha_{\rm max}\!\ll\!1$, we obtain $\Delta{\tilde L}(t)/\Delta{x}_{\rm max}(t)\!\ll\!1$ as an upper bound, where $\Delta{\tilde L}(t)\={\tilde L}(t)\!-\!{\tilde L}(t_0)$ and $\Delta{x}_{\rm max}(t)\=x_{\rm{max}}(t)-x_{\rm{max}}(t_0)$ (here, $t_0$ is the first time point of each growing pulse presented in Fig.~\ref{fig:fig3}).

The integral corollary $\Delta{\tilde L}(t)/\Delta{x}_{\rm max}(t)\!\ll\!1$ of Eq.~\eqref{eq:slowness} in the manuscript is satisfied for all growing pulses observed in our experiments, where two representative examples are shown in Fig.~\ref{fig:fig4} in the manuscript. It is important to stress that $\Delta{\tilde L}(t)/\Delta{x}_{\rm max}(t)$ is an upper bound on $\dot{L}(t)/c_{\rm r}(t)$. Consequently, the latter features smaller values than those observed in Fig.~\ref{fig:fig4} in the manuscript, in line with the theoretical prediction in Eq.~\eqref{eq:slowness} in the manuscript.



\begin{thebibliography}{70}%
\makeatletter
\providecommand \@ifxundefined [1]{%
 \@ifx{#1\undefined}
}%
\providecommand \@ifnum [1]{%
 \ifnum #1\expandafter \@firstoftwo
 \else \expandafter \@secondoftwo
 \fi
}%
\providecommand \@ifx [1]{%
 \ifx #1\expandafter \@firstoftwo
 \else \expandafter \@secondoftwo
 \fi
}%
\providecommand \natexlab [1]{#1}%
\providecommand \enquote  [1]{``#1''}%
\providecommand \bibnamefont  [1]{#1}%
\providecommand \bibfnamefont [1]{#1}%
\providecommand \citenamefont [1]{#1}%
\providecommand \href@noop [0]{\@secondoftwo}%
\providecommand \href [0]{\begingroup \@sanitize@url \@href}%
\providecommand \@href[1]{\@@startlink{#1}\@@href}%
\providecommand \@@href[1]{\endgroup#1\@@endlink}%
\providecommand \@sanitize@url [0]{\catcode `\\12\catcode `\$12\catcode
  `\&12\catcode `\#12\catcode `\^12\catcode `\_12\catcode `\%12\relax}%
\providecommand \@@startlink[1]{}%
\providecommand \@@endlink[0]{}%
\providecommand \url  [0]{\begingroup\@sanitize@url \@url }%
\providecommand \@url [1]{\endgroup\@href {#1}{\urlprefix }}%
\providecommand \urlprefix  [0]{URL }%
\providecommand \Eprint [0]{\href }%
\providecommand \doibase [0]{https://doi.org/}%
\providecommand \selectlanguage [0]{\@gobble}%
\providecommand \bibinfo  [0]{\@secondoftwo}%
\providecommand \bibfield  [0]{\@secondoftwo}%
\providecommand \translation [1]{[#1]}%
\providecommand \BibitemOpen [0]{}%
\providecommand \bibitemStop [0]{}%
\providecommand \bibitemNoStop [0]{.\EOS\space}%
\providecommand \EOS [0]{\spacefactor3000\relax}%
\providecommand \BibitemShut  [1]{\csname bibitem#1\endcsname}%
\let\auto@bib@innerbib\@empty
\bibitem [{\citenamefont {Ben-Zion}(2001)}]{Ben-Zion2001}%
  \BibitemOpen
  \bibfield  {author} {\bibinfo {author} {\bibfnamefont {Y.}~\bibnamefont
  {Ben-Zion}},\ }\bibfield  {title} {\bibinfo {title} {{Dynamic ruptures in
  recent models of earthquake faults}},\ }\href
  {https://doi.org/10.1016/S0022-5096(01)00036-9} {\bibfield  {journal}
  {\bibinfo  {journal} {J. Mech. Phys. Solids}\ }\textbf {\bibinfo {volume}
  {49}},\ \bibinfo {pages} {2209} (\bibinfo {year} {2001})}\BibitemShut
  {NoStop}%
\bibitem [{\citenamefont {Scholz}(2002)}]{Scholz2002}%
  \BibitemOpen
  \bibfield  {author} {\bibinfo {author} {\bibfnamefont {C.~H.}\ \bibnamefont
  {Scholz}},\ }\href {https://doi.org/10.1017/9781316681473} {\emph {\bibinfo
  {title} {{The Mechanics of Earthquakes and Faulting}}}}\ (\bibinfo
  {publisher} {Cambridge university press},\ \bibinfo {year}
  {2002})\BibitemShut {NoStop}%
\bibitem [{\citenamefont {Lu}\ \emph {et~al.}(2007)\citenamefont {Lu},
  \citenamefont {Lapusta},\ and\ \citenamefont {Rosakis}}]{Lu2007}%
  \BibitemOpen
  \bibfield  {author} {\bibinfo {author} {\bibfnamefont {X.}~\bibnamefont
  {Lu}}, \bibinfo {author} {\bibfnamefont {N.}~\bibnamefont {Lapusta}},\ and\
  \bibinfo {author} {\bibfnamefont {A.~J.}\ \bibnamefont {Rosakis}},\
  }\bibfield  {title} {\bibinfo {title} {Pulse-like and crack-like ruptures in
  experiments mimicking crustal earthquakes},\ }\href
  {https://doi.org/10.1073/pnas.0704268104} {\bibfield  {journal} {\bibinfo
  {journal} {Proc. Natl. Acad. Sci.}\ }\textbf {\bibinfo {volume} {104}},\
  \bibinfo {pages} {18931} (\bibinfo {year} {2007})}\BibitemShut {NoStop}%
\bibitem [{\citenamefont {Lu}\ \emph {et~al.}(2010)\citenamefont {Lu},
  \citenamefont {Lapusta},\ and\ \citenamefont {Rosakis}}]{Lu2010a}%
  \BibitemOpen
  \bibfield  {author} {\bibinfo {author} {\bibfnamefont {X.}~\bibnamefont
  {Lu}}, \bibinfo {author} {\bibfnamefont {N.}~\bibnamefont {Lapusta}},\ and\
  \bibinfo {author} {\bibfnamefont {A.~J.}\ \bibnamefont {Rosakis}},\
  }\bibfield  {title} {\bibinfo {title} {{Pulse-like and crack-like dynamic
  shear ruptures on frictional interfaces: experimental evidence, numerical
  modeling, and implications}},\ }\href
  {https://doi.org/10.1007/s10704-010-9479-4} {\bibfield  {journal} {\bibinfo
  {journal} {Int. J. Fract.}\ }\textbf {\bibinfo {volume} {163}},\ \bibinfo
  {pages} {27} (\bibinfo {year} {2010})}\BibitemShut {NoStop}%
\bibitem [{\citenamefont {Svetlizky}\ \emph {et~al.}(2019)\citenamefont
  {Svetlizky}, \citenamefont {Bayart},\ and\ \citenamefont
  {Fineberg}}]{Svetlizky2019}%
  \BibitemOpen
  \bibfield  {author} {\bibinfo {author} {\bibfnamefont {I.}~\bibnamefont
  {Svetlizky}}, \bibinfo {author} {\bibfnamefont {E.}~\bibnamefont {Bayart}},\
  and\ \bibinfo {author} {\bibfnamefont {J.}~\bibnamefont {Fineberg}},\
  }\bibfield  {title} {\bibinfo {title} {{Brittle Fracture Theory Describes the
  Onset of Frictional Motion}},\ }\href
  {https://doi.org/10.1146/annurev-conmatphys-031218-013327} {\bibfield
  {journal} {\bibinfo  {journal} {Annu. Rev. Condens. Matter Phys.}\ }\textbf
  {\bibinfo {volume} {10}},\ \bibinfo {pages} {031218} (\bibinfo {year}
  {2019})}\BibitemShut {NoStop}%
\bibitem [{\citenamefont {Freund}(1979)}]{freund1979mechanics}%
  \BibitemOpen
  \bibfield  {author} {\bibinfo {author} {\bibfnamefont {L.~B.}\ \bibnamefont
  {Freund}},\ }\bibfield  {title} {\bibinfo {title} {The mechanics of dynamic
  shear crack propagation},\ }\href
  {https://agupubs.onlinelibrary.wiley.com/doi/abs/10.1029/JB084iB05p02199}
  {\bibfield  {journal} {\bibinfo  {journal} {J. Geophys. Res. Solid Earth}\
  }\textbf {\bibinfo {volume} {84}},\ \bibinfo {pages} {2199} (\bibinfo {year}
  {1979})}\BibitemShut {NoStop}%
\bibitem [{\citenamefont {Heaton}(1990)}]{Heaton1990}%
  \BibitemOpen
  \bibfield  {author} {\bibinfo {author} {\bibfnamefont {T.~H.}\ \bibnamefont
  {Heaton}},\ }\bibfield  {title} {\bibinfo {title} {{Evidence for and
  implications of self-healing pulses of slip in earthquake rupture}},\ }\href
  {https://doi.org/10.1016/0031-9201(90)90002-F} {\bibfield  {journal}
  {\bibinfo  {journal} {Phys. Earth Planet. Inter.}\ }\textbf {\bibinfo
  {volume} {64}},\ \bibinfo {pages} {1} (\bibinfo {year} {1990})}\BibitemShut
  {NoStop}%
\bibitem [{\citenamefont {Perrin}\ \emph {et~al.}(1995)\citenamefont {Perrin},
  \citenamefont {Rice},\ and\ \citenamefont {Zheng}}]{Perrin1995}%
  \BibitemOpen
  \bibfield  {author} {\bibinfo {author} {\bibfnamefont {G.}~\bibnamefont
  {Perrin}}, \bibinfo {author} {\bibfnamefont {J.~R.}\ \bibnamefont {Rice}},\
  and\ \bibinfo {author} {\bibfnamefont {G.}~\bibnamefont {Zheng}},\ }\bibfield
   {title} {\bibinfo {title} {{Self-healing slip pulse on a frictional
  surface}},\ }\href {https://doi.org/10.1016/0022-5096(95)00036-I} {\bibfield
  {journal} {\bibinfo  {journal} {J. Mech. Phys. Solids}\ }\textbf {\bibinfo
  {volume} {43}},\ \bibinfo {pages} {1461} (\bibinfo {year}
  {1995})}\BibitemShut {NoStop}%
\bibitem [{\citenamefont {Beroza}\ and\ \citenamefont
  {Mikumo}(1996)}]{Beroza_Mikumo_1996}%
  \BibitemOpen
  \bibfield  {author} {\bibinfo {author} {\bibfnamefont {G.~C.}\ \bibnamefont
  {Beroza}}\ and\ \bibinfo {author} {\bibfnamefont {T.}~\bibnamefont
  {Mikumo}},\ }\bibfield  {title} {\bibinfo {title} {Short slip duration in
  dynamic rupture in the presence of heterogeneous fault properties},\ }\href
  {https://doi.org/https://doi.org/10.1029/96JB02291} {\bibfield  {journal}
  {\bibinfo  {journal} {J. Geophys. Res. Solid Earth}\ }\textbf {\bibinfo
  {volume} {101}},\ \bibinfo {pages} {22449} (\bibinfo {year}
  {1996})}\BibitemShut {NoStop}%
\bibitem [{\citenamefont {Beeler}\ and\ \citenamefont
  {Tullis}(1996)}]{Beeler1996}%
  \BibitemOpen
  \bibfield  {author} {\bibinfo {author} {\bibfnamefont {N.~M.}\ \bibnamefont
  {Beeler}}\ and\ \bibinfo {author} {\bibfnamefont {T.~E.}\ \bibnamefont
  {Tullis}},\ }\bibfield  {title} {\bibinfo {title} {{Self-healing slip pulses
  in dynamic rupture models due to velocity-dependent strength}},\ }\href
  {https://pubs.geoscienceworld.org/ssa/bssa/article/86/4/1130/120159/self-healing-slip-pulses-in-dynamic-rupture-models}
  {\bibfield  {journal} {\bibinfo  {journal} {Bull. Seismol. Soc. Am.}\
  }\textbf {\bibinfo {volume} {86}},\ \bibinfo {pages} {1130} (\bibinfo {year}
  {1996})}\BibitemShut {NoStop}%
\bibitem [{\citenamefont {Cochard}\ and\ \citenamefont
  {Madariaga}(1996)}]{Cochard1996}%
  \BibitemOpen
  \bibfield  {author} {\bibinfo {author} {\bibfnamefont {A.}~\bibnamefont
  {Cochard}}\ and\ \bibinfo {author} {\bibfnamefont {R.}~\bibnamefont
  {Madariaga}},\ }\bibfield  {title} {\bibinfo {title} {{Complexity of
  seismicity due to highly rate-dependent friction}},\ }\href
  {https://doi.org/10.1029/96JB02095} {\bibfield  {journal} {\bibinfo
  {journal} {J. Geophys. Res. Solid Earth}\ }\textbf {\bibinfo {volume}
  {101}},\ \bibinfo {pages} {25321} (\bibinfo {year} {1996})}\BibitemShut
  {NoStop}%
\bibitem [{\citenamefont {Andrews}\ and\ \citenamefont
  {Ben-Zion}(1997)}]{Andrews1997}%
  \BibitemOpen
  \bibfield  {author} {\bibinfo {author} {\bibfnamefont {D.~J.}\ \bibnamefont
  {Andrews}}\ and\ \bibinfo {author} {\bibfnamefont {Y.}~\bibnamefont
  {Ben-Zion}},\ }\bibfield  {title} {\bibinfo {title} {{Wrinkle-like slip pulse
  on a fault between different materials}},\ }\href
  {https://doi.org/10.1029/96JB02856} {\bibfield  {journal} {\bibinfo
  {journal} {J. Geophys. Res. Solid Earth}\ }\textbf {\bibinfo {volume}
  {102}},\ \bibinfo {pages} {553} (\bibinfo {year} {1997})}\BibitemShut
  {NoStop}%
\bibitem [{\citenamefont {Zheng}\ and\ \citenamefont {Rice}(1998)}]{Zheng1998}%
  \BibitemOpen
  \bibfield  {author} {\bibinfo {author} {\bibfnamefont {G.}~\bibnamefont
  {Zheng}}\ and\ \bibinfo {author} {\bibfnamefont {J.~R.}\ \bibnamefont
  {Rice}},\ }\bibfield  {title} {\bibinfo {title} {{Conditions under which
  velocity-weakening friction allows a self-healing versus a cracklike mode of
  rupture}},\ }\href
  {https://pubs.geoscienceworld.org/ssa/bssa/article/88/6/1466/120398/conditions-under-which-velocity-weakening-friction}
  {\bibfield  {journal} {\bibinfo  {journal} {Bull. Seismol. Soc. Am.}\
  }\textbf {\bibinfo {volume} {88}},\ \bibinfo {pages} {1466} (\bibinfo {year}
  {1998})}\BibitemShut {NoStop}%
\bibitem [{\citenamefont {Nielsen}\ \emph {et~al.}(2000)\citenamefont
  {Nielsen}, \citenamefont {Carlson},\ and\ \citenamefont
  {Olsen}}]{Nielsen2000}%
  \BibitemOpen
  \bibfield  {author} {\bibinfo {author} {\bibfnamefont {S.~B.}\ \bibnamefont
  {Nielsen}}, \bibinfo {author} {\bibfnamefont {J.~M.}\ \bibnamefont
  {Carlson}},\ and\ \bibinfo {author} {\bibfnamefont {K.~B.}\ \bibnamefont
  {Olsen}},\ }\bibfield  {title} {\bibinfo {title} {{Influence of friction and
  fault geometry on earthquake rupture}},\ }\href
  {https://doi.org/10.1029/1999JB900350} {\bibfield  {journal} {\bibinfo
  {journal} {J. Geophys. Res. Solid Earth}\ }\textbf {\bibinfo {volume}
  {105}},\ \bibinfo {pages} {6069} (\bibinfo {year} {2000})}\BibitemShut
  {NoStop}%
\bibitem [{\citenamefont {Nielsen}\ and\ \citenamefont
  {Madariaga}(2003)}]{Nielsen2003}%
  \BibitemOpen
  \bibfield  {author} {\bibinfo {author} {\bibfnamefont {S.}~\bibnamefont
  {Nielsen}}\ and\ \bibinfo {author} {\bibfnamefont {R.}~\bibnamefont
  {Madariaga}},\ }\bibfield  {title} {\bibinfo {title} {{On the Self-Healing
  Fracture Mode}},\ }\href {https://doi.org/10.1785/0120020090} {\bibfield
  {journal} {\bibinfo  {journal} {Bull. Seismol. Soc. Am.}\ }\textbf {\bibinfo
  {volume} {93}},\ \bibinfo {pages} {2375} (\bibinfo {year}
  {2003})}\BibitemShut {NoStop}%
\bibitem [{\citenamefont {Brener}\ \emph {et~al.}(2005)\citenamefont {Brener},
  \citenamefont {Malinin},\ and\ \citenamefont {Marchenko}}]{Brener2005}%
  \BibitemOpen
  \bibfield  {author} {\bibinfo {author} {\bibfnamefont {E.~A.}\ \bibnamefont
  {Brener}}, \bibinfo {author} {\bibfnamefont {S.~V.}\ \bibnamefont
  {Malinin}},\ and\ \bibinfo {author} {\bibfnamefont {V.~I.}\ \bibnamefont
  {Marchenko}},\ }\bibfield  {title} {\bibinfo {title} {{Fracture and friction:
  Stick-slip motion}},\ }\href {https://doi.org/10.1140/epje/i2004-10112-3}
  {\bibfield  {journal} {\bibinfo  {journal} {Eur. Phys. J. E}\ }\textbf
  {\bibinfo {volume} {17}},\ \bibinfo {pages} {101} (\bibinfo {year}
  {2005})}\BibitemShut {NoStop}%
\bibitem [{\citenamefont {Dunham}\ and\ \citenamefont
  {Archuleta}(2005)}]{Dunham2005}%
  \BibitemOpen
  \bibfield  {author} {\bibinfo {author} {\bibfnamefont {E.~M.}\ \bibnamefont
  {Dunham}}\ and\ \bibinfo {author} {\bibfnamefont {R.~J.}\ \bibnamefont
  {Archuleta}},\ }\bibfield  {title} {\bibinfo {title} {{Near-source ground
  motion from steady state dynamic rupture pulses}},\ }\href
  {https://doi.org/10.1029/2004GL021793} {\bibfield  {journal} {\bibinfo
  {journal} {Geophys. Res. Lett.}\ }\textbf {\bibinfo {volume} {32}},\ \bibinfo
  {pages} {L03302} (\bibinfo {year} {2005})}\BibitemShut {NoStop}%
\bibitem [{\citenamefont {Lykotrafitis}\ \emph {et~al.}(2006)\citenamefont
  {Lykotrafitis}, \citenamefont {Rosakis},\ and\ \citenamefont
  {Ravichandran}}]{lykotrafitis2006self}%
  \BibitemOpen
  \bibfield  {author} {\bibinfo {author} {\bibfnamefont {G.}~\bibnamefont
  {Lykotrafitis}}, \bibinfo {author} {\bibfnamefont {A.~J.}\ \bibnamefont
  {Rosakis}},\ and\ \bibinfo {author} {\bibfnamefont {G.}~\bibnamefont
  {Ravichandran}},\ }\bibfield  {title} {\bibinfo {title} {Self-healing
  pulse-like shear ruptures in the laboratory},\ }\href
  {https://doi.org/10.1126/science.1128359} {\bibfield  {journal} {\bibinfo
  {journal} {Science}\ }\textbf {\bibinfo {volume} {313}},\ \bibinfo {pages}
  {1765} (\bibinfo {year} {2006})}\BibitemShut {NoStop}%
\bibitem [{\citenamefont {Shi}\ \emph {et~al.}(2008)\citenamefont {Shi},
  \citenamefont {Ben-Zion},\ and\ \citenamefont {Needleman}}]{Shi2008}%
  \BibitemOpen
  \bibfield  {author} {\bibinfo {author} {\bibfnamefont {Z.}~\bibnamefont
  {Shi}}, \bibinfo {author} {\bibfnamefont {Y.}~\bibnamefont {Ben-Zion}},\ and\
  \bibinfo {author} {\bibfnamefont {A.}~\bibnamefont {Needleman}},\ }\bibfield
  {title} {\bibinfo {title} {{Properties of dynamic rupture and energy
  partition in a solid with a frictional interface}},\ }\href
  {https://doi.org/10.1016/j.jmps.2007.04.006} {\bibfield  {journal} {\bibinfo
  {journal} {J. Mech. Phys. Solids}\ }\textbf {\bibinfo {volume} {56}},\
  \bibinfo {pages} {5} (\bibinfo {year} {2008})}\BibitemShut {NoStop}%
\bibitem [{\citenamefont {Rubin}\ and\ \citenamefont
  {Ampuero}(2009)}]{Rubin2009}%
  \BibitemOpen
  \bibfield  {author} {\bibinfo {author} {\bibfnamefont {A.~M.}\ \bibnamefont
  {Rubin}}\ and\ \bibinfo {author} {\bibfnamefont {J.-P.}\ \bibnamefont
  {Ampuero}},\ }\bibfield  {title} {\bibinfo {title} {{Self-similar slip pulses
  during rate-and-state earthquake nucleation}},\ }\href
  {https://doi.org/10.1029/2009JB006529} {\bibfield  {journal} {\bibinfo
  {journal} {J. Geophys. Res.}\ }\textbf {\bibinfo {volume} {114}},\ \bibinfo
  {pages} {B11305} (\bibinfo {year} {2009})}\BibitemShut {NoStop}%
\bibitem [{\citenamefont {Garagash}(2012)}]{Garagash2012}%
  \BibitemOpen
  \bibfield  {author} {\bibinfo {author} {\bibfnamefont {D.~I.}\ \bibnamefont
  {Garagash}},\ }\bibfield  {title} {\bibinfo {title} {{Seismic and aseismic
  slip pulses driven by thermal pressurization of pore fluid}},\ }\href
  {https://doi.org/10.1029/2011JB008889} {\bibfield  {journal} {\bibinfo
  {journal} {J. Geophys. Res. Solid Earth}\ }\textbf {\bibinfo {volume}
  {117}},\ \bibinfo {pages} {B04314} (\bibinfo {year} {2012})}\BibitemShut
  {NoStop}%
\bibitem [{\citenamefont {Platt}\ \emph {et~al.}(2015)\citenamefont {Platt},
  \citenamefont {Viesca},\ and\ \citenamefont {Garagash}}]{platt2015steadily}%
  \BibitemOpen
  \bibfield  {author} {\bibinfo {author} {\bibfnamefont {J.~D.}\ \bibnamefont
  {Platt}}, \bibinfo {author} {\bibfnamefont {R.~C.}\ \bibnamefont {Viesca}},\
  and\ \bibinfo {author} {\bibfnamefont {D.~I.}\ \bibnamefont {Garagash}},\
  }\bibfield  {title} {\bibinfo {title} {Steadily propagating slip pulses
  driven by thermal decomposition},\ }\href
  {https://doi.org/10.1002/2015JB012200} {\bibfield  {journal} {\bibinfo
  {journal} {Journal of Geophysical Research: Solid Earth}\ }\textbf {\bibinfo
  {volume} {120}},\ \bibinfo {pages} {6558} (\bibinfo {year}
  {2015})}\BibitemShut {NoStop}%
\bibitem [{\citenamefont {Gabriel}\ \emph {et~al.}(2012)\citenamefont
  {Gabriel}, \citenamefont {Ampuero}, \citenamefont {Dalguer},\ and\
  \citenamefont {Mai}}]{Gabriel2012}%
  \BibitemOpen
  \bibfield  {author} {\bibinfo {author} {\bibfnamefont {A.-A.}\ \bibnamefont
  {Gabriel}}, \bibinfo {author} {\bibfnamefont {J.-P.}\ \bibnamefont
  {Ampuero}}, \bibinfo {author} {\bibfnamefont {L.~A.}\ \bibnamefont
  {Dalguer}},\ and\ \bibinfo {author} {\bibfnamefont {P.~M.}\ \bibnamefont
  {Mai}},\ }\bibfield  {title} {\bibinfo {title} {{The transition of dynamic
  rupture styles in elastic media under velocity-weakening friction}},\ }\href
  {https://doi.org/10.1029/2012JB009468} {\bibfield  {journal} {\bibinfo
  {journal} {J. Geophys. Res. Solid Earth}\ }\textbf {\bibinfo {volume}
  {117}},\ \bibinfo {pages} {B09311} (\bibinfo {year} {2012})}\BibitemShut
  {NoStop}%
\bibitem [{\citenamefont {Putelat}\ \emph {et~al.}(2017)\citenamefont
  {Putelat}, \citenamefont {Dawes},\ and\ \citenamefont
  {Champneys}}]{Putelat2017}%
  \BibitemOpen
  \bibfield  {author} {\bibinfo {author} {\bibfnamefont {T.}~\bibnamefont
  {Putelat}}, \bibinfo {author} {\bibfnamefont {J.~H.}\ \bibnamefont {Dawes}},\
  and\ \bibinfo {author} {\bibfnamefont {A.~R.}\ \bibnamefont {Champneys}},\
  }\bibfield  {title} {\bibinfo {title} {{A phase-plane analysis of localized
  frictional waves}},\ }\href {https://doi.org/10.1098/rspa.2016.0606}
  {\bibfield  {journal} {\bibinfo  {journal} {Proc. R. Soc. A Math. Phys. Eng.
  Sci.}\ }\textbf {\bibinfo {volume} {473}},\ \bibinfo {pages} {20160606}
  (\bibinfo {year} {2017})}\BibitemShut {NoStop}%
\bibitem [{\citenamefont {Michel}\ \emph {et~al.}(2017)\citenamefont {Michel},
  \citenamefont {Avouac}, \citenamefont {Lapusta},\ and\ \citenamefont
  {Jiang}}]{Michel2017}%
  \BibitemOpen
  \bibfield  {author} {\bibinfo {author} {\bibfnamefont {S.}~\bibnamefont
  {Michel}}, \bibinfo {author} {\bibfnamefont {J.-P.}\ \bibnamefont {Avouac}},
  \bibinfo {author} {\bibfnamefont {N.}~\bibnamefont {Lapusta}},\ and\ \bibinfo
  {author} {\bibfnamefont {J.}~\bibnamefont {Jiang}},\ }\bibfield  {title}
  {\bibinfo {title} {{Pulse-like partial ruptures and high-frequency radiation
  at creeping-locked transition during megathrust earthquakes}},\ }\href
  {https://doi.org/10.1002/2017GL074725} {\bibfield  {journal} {\bibinfo
  {journal} {Geophys. Res. Lett.}\ }\textbf {\bibinfo {volume} {44}},\ \bibinfo
  {pages} {8345} (\bibinfo {year} {2017})}\BibitemShut {NoStop}%
\bibitem [{\citenamefont {Kohrangi}\ \emph {et~al.}(2019)\citenamefont
  {Kohrangi}, \citenamefont {Vamvatsikos},\ and\ \citenamefont
  {Bazzurro}}]{kohrangi2019pulse}%
  \BibitemOpen
  \bibfield  {author} {\bibinfo {author} {\bibfnamefont {M.}~\bibnamefont
  {Kohrangi}}, \bibinfo {author} {\bibfnamefont {D.}~\bibnamefont
  {Vamvatsikos}},\ and\ \bibinfo {author} {\bibfnamefont {P.}~\bibnamefont
  {Bazzurro}},\ }\bibfield  {title} {\bibinfo {title} {Pulse-like versus
  non-pulse-like ground motion records: spectral shape comparisons and record
  selection strategies},\ }\href {https://doi.org/10.1002/eqe.3122} {\bibfield
  {journal} {\bibinfo  {journal} {Earthquake Engineering \& Structural
  Dynamics}\ }\textbf {\bibinfo {volume} {48}},\ \bibinfo {pages} {46}
  (\bibinfo {year} {2019})}\BibitemShut {NoStop}%
\bibitem [{\citenamefont {Chen}\ \emph {et~al.}(2020)\citenamefont {Chen},
  \citenamefont {Avouac}, \citenamefont {Aati}, \citenamefont {Milliner},
  \citenamefont {Zheng},\ and\ \citenamefont {Shi}}]{chen2020cascading}%
  \BibitemOpen
  \bibfield  {author} {\bibinfo {author} {\bibfnamefont {K.}~\bibnamefont
  {Chen}}, \bibinfo {author} {\bibfnamefont {J.-P.}\ \bibnamefont {Avouac}},
  \bibinfo {author} {\bibfnamefont {S.}~\bibnamefont {Aati}}, \bibinfo {author}
  {\bibfnamefont {C.}~\bibnamefont {Milliner}}, \bibinfo {author}
  {\bibfnamefont {F.}~\bibnamefont {Zheng}},\ and\ \bibinfo {author}
  {\bibfnamefont {C.}~\bibnamefont {Shi}},\ }\bibfield  {title} {\bibinfo
  {title} {Cascading and pulse-like ruptures during the 2019 ridgecrest
  earthquakes in the eastern california shear zone},\ }\href
  {https://doi.org/10.1038/s41467-019-13750-w} {\bibfield  {journal} {\bibinfo
  {journal} {Nature Communications}\ }\textbf {\bibinfo {volume} {11}},\
  \bibinfo {pages} {22} (\bibinfo {year} {2020})}\BibitemShut {NoStop}%
\bibitem [{\citenamefont {Wu}\ \emph {et~al.}(2023)\citenamefont {Wu},
  \citenamefont {Xie}, \citenamefont {An}, \citenamefont {Lyu}, \citenamefont
  {Taymaz}, \citenamefont {Irmak}, \citenamefont {Li}, \citenamefont {Wen},\
  and\ \citenamefont {Zhou}}]{wu2023pulse}%
  \BibitemOpen
  \bibfield  {author} {\bibinfo {author} {\bibfnamefont {F.}~\bibnamefont
  {Wu}}, \bibinfo {author} {\bibfnamefont {J.}~\bibnamefont {Xie}}, \bibinfo
  {author} {\bibfnamefont {Z.}~\bibnamefont {An}}, \bibinfo {author}
  {\bibfnamefont {C.}~\bibnamefont {Lyu}}, \bibinfo {author} {\bibfnamefont
  {T.}~\bibnamefont {Taymaz}}, \bibinfo {author} {\bibfnamefont {T.~S.}\
  \bibnamefont {Irmak}}, \bibinfo {author} {\bibfnamefont {X.}~\bibnamefont
  {Li}}, \bibinfo {author} {\bibfnamefont {Z.}~\bibnamefont {Wen}},\ and\
  \bibinfo {author} {\bibfnamefont {B.}~\bibnamefont {Zhou}},\ }\bibfield
  {title} {\bibinfo {title} {{Pulse-like ground motion observed during the 6
  February 2023 M$_{\rm W}$7.8 Pazarc{\i}k Earthquake (Kahramanmara{\c{s}}, SE
  T{\"u}rkiye)}},\ }\href {https://doi.org/10.1016/j.eqs.2023.05.005}
  {\bibfield  {journal} {\bibinfo  {journal} {Earthquake Science}\ }\textbf
  {\bibinfo {volume} {36}},\ \bibinfo {pages} {328} (\bibinfo {year}
  {2023})}\BibitemShut {NoStop}%
\bibitem [{\citenamefont {Roch}\ \emph {et~al.}(2022)\citenamefont {Roch},
  \citenamefont {Brener}, \citenamefont {Molinari},\ and\ \citenamefont
  {Bouchbinder}}]{ROCH2022104607}%
  \BibitemOpen
  \bibfield  {author} {\bibinfo {author} {\bibfnamefont {T.}~\bibnamefont
  {Roch}}, \bibinfo {author} {\bibfnamefont {E.~A.}\ \bibnamefont {Brener}},
  \bibinfo {author} {\bibfnamefont {J.-F.}\ \bibnamefont {Molinari}},\ and\
  \bibinfo {author} {\bibfnamefont {E.}~\bibnamefont {Bouchbinder}},\
  }\bibfield  {title} {\bibinfo {title} {Velocity-driven frictional sliding:
  Coarsening and steady-state pulses},\ }\href
  {https://doi.org/https://doi.org/10.1016/j.jmps.2021.104607} {\bibfield
  {journal} {\bibinfo  {journal} {J. Mech. Phys. Solids}\ }\textbf {\bibinfo
  {volume} {158}},\ \bibinfo {pages} {104607} (\bibinfo {year}
  {2022})}\BibitemShut {NoStop}%
\bibitem [{\citenamefont {Galetzka}\ \emph {et~al.}(2015)\citenamefont
  {Galetzka}, \citenamefont {Melgar}, \citenamefont {Genrich}, \citenamefont
  {Geng}, \citenamefont {Owen}, \citenamefont {Lindsey}, \citenamefont {Xu},
  \citenamefont {Bock}, \citenamefont {Avouac}, \citenamefont {Adhikari} \emph
  {et~al.}}]{galetzka2015slip}%
  \BibitemOpen
  \bibfield  {author} {\bibinfo {author} {\bibfnamefont {J.}~\bibnamefont
  {Galetzka}}, \bibinfo {author} {\bibfnamefont {D.}~\bibnamefont {Melgar}},
  \bibinfo {author} {\bibfnamefont {J.~F.}\ \bibnamefont {Genrich}}, \bibinfo
  {author} {\bibfnamefont {J.}~\bibnamefont {Geng}}, \bibinfo {author}
  {\bibfnamefont {S.}~\bibnamefont {Owen}}, \bibinfo {author} {\bibfnamefont
  {E.}~\bibnamefont {Lindsey}}, \bibinfo {author} {\bibfnamefont
  {X.}~\bibnamefont {Xu}}, \bibinfo {author} {\bibfnamefont {Y.}~\bibnamefont
  {Bock}}, \bibinfo {author} {\bibfnamefont {J.-P.}\ \bibnamefont {Avouac}},
  \bibinfo {author} {\bibfnamefont {L.}~\bibnamefont {Adhikari}}, \emph
  {et~al.},\ }\bibfield  {title} {\bibinfo {title} {Slip pulse and resonance of
  the kathmandu basin during the 2015 gorkha earthquake, nepal},\ }\href
  {https://doi.org/10.1126/science.aac6383} {\bibfield  {journal} {\bibinfo
  {journal} {Science}\ }\textbf {\bibinfo {volume} {349}},\ \bibinfo {pages}
  {1091} (\bibinfo {year} {2015})}\BibitemShut {NoStop}%
\bibitem [{\citenamefont {Dunham}\ \emph {et~al.}(2011)\citenamefont {Dunham},
  \citenamefont {Belanger}, \citenamefont {Cong},\ and\ \citenamefont
  {Kozdon}}]{Dunham2011a}%
  \BibitemOpen
  \bibfield  {author} {\bibinfo {author} {\bibfnamefont {E.~M.}\ \bibnamefont
  {Dunham}}, \bibinfo {author} {\bibfnamefont {D.}~\bibnamefont {Belanger}},
  \bibinfo {author} {\bibfnamefont {L.}~\bibnamefont {Cong}},\ and\ \bibinfo
  {author} {\bibfnamefont {J.~E.}\ \bibnamefont {Kozdon}},\ }\bibfield  {title}
  {\bibinfo {title} {Earthquake ruptures with strongly rate-weakening friction
  and off-fault plasticity, part 1: Planar faults},\ }\href
  {https://doi.org/10.1785/0120100075} {\bibfield  {journal} {\bibinfo
  {journal} {Bulletin of the Seismological Society of America}\ }\textbf
  {\bibinfo {volume} {101}},\ \bibinfo {pages} {2296} (\bibinfo {year}
  {2011})}\BibitemShut {NoStop}%
\bibitem [{\citenamefont {Brener}\ \emph {et~al.}(2018)\citenamefont {Brener},
  \citenamefont {Aldam}, \citenamefont {Barras}, \citenamefont {Molinari},\
  and\ \citenamefont {Bouchbinder}}]{Brener2018}%
  \BibitemOpen
  \bibfield  {author} {\bibinfo {author} {\bibfnamefont {E.~A.}\ \bibnamefont
  {Brener}}, \bibinfo {author} {\bibfnamefont {M.}~\bibnamefont {Aldam}},
  \bibinfo {author} {\bibfnamefont {F.}~\bibnamefont {Barras}}, \bibinfo
  {author} {\bibfnamefont {J.-F.}\ \bibnamefont {Molinari}},\ and\ \bibinfo
  {author} {\bibfnamefont {E.}~\bibnamefont {Bouchbinder}},\ }\bibfield
  {title} {\bibinfo {title} {{Unstable Slip Pulses and Earthquake Nucleation as
  a Nonequilibrium First-Order Phase Transition}},\ }\href
  {https://doi.org/10.1103/PhysRevLett.121.234302} {\bibfield  {journal}
  {\bibinfo  {journal} {Phys. Rev. Lett.}\ }\textbf {\bibinfo {volume} {121}},\
  \bibinfo {pages} {234302} (\bibinfo {year} {2018})}\BibitemShut {NoStop}%
\bibitem [{\citenamefont {Pomyalov}\ \emph
  {et~al.}(2023{\natexlab{a}})\citenamefont {Pomyalov}, \citenamefont
  {Lubomirsky}, \citenamefont {Braverman}, \citenamefont {Brener},\ and\
  \citenamefont {Bouchbinder}}]{pomyalov2023self}%
  \BibitemOpen
  \bibfield  {author} {\bibinfo {author} {\bibfnamefont {A.}~\bibnamefont
  {Pomyalov}}, \bibinfo {author} {\bibfnamefont {Y.}~\bibnamefont
  {Lubomirsky}}, \bibinfo {author} {\bibfnamefont {L.}~\bibnamefont
  {Braverman}}, \bibinfo {author} {\bibfnamefont {E.~A.}\ \bibnamefont
  {Brener}},\ and\ \bibinfo {author} {\bibfnamefont {E.}~\bibnamefont
  {Bouchbinder}},\ }\bibfield  {title} {\bibinfo {title} {Self-healing
  solitonic slip pulses in frictional systems},\ }\href
  {https://doi.org/10.1103/PhysRevE.107.L013001} {\bibfield  {journal}
  {\bibinfo  {journal} {Physical Review E}\ }\textbf {\bibinfo {volume}
  {107}},\ \bibinfo {pages} {L013001} (\bibinfo {year}
  {2023}{\natexlab{a}})}\BibitemShut {NoStop}%
\bibitem [{\citenamefont {Brantut}\ \emph {et~al.}(2019)\citenamefont
  {Brantut}, \citenamefont {Garagash},\ and\ \citenamefont
  {Noda}}]{brantut2019stability}%
  \BibitemOpen
  \bibfield  {author} {\bibinfo {author} {\bibfnamefont {N.}~\bibnamefont
  {Brantut}}, \bibinfo {author} {\bibfnamefont {D.~I.}\ \bibnamefont
  {Garagash}},\ and\ \bibinfo {author} {\bibfnamefont {H.}~\bibnamefont
  {Noda}},\ }\bibfield  {title} {\bibinfo {title} {Stability of pulse-like
  earthquake ruptures},\ }\href
  {https://agupubs.onlinelibrary.wiley.com/doi/abs/10.1029/2019JB017926}
  {\bibfield  {journal} {\bibinfo  {journal} {J. Geophys. Res. Solid Earth}\
  }\textbf {\bibinfo {volume} {124}},\ \bibinfo {pages} {8998} (\bibinfo {year}
  {2019})}\BibitemShut {NoStop}%
\bibitem [{\citenamefont {Pomyalov}\ \emph
  {et~al.}(2023{\natexlab{b}})\citenamefont {Pomyalov}, \citenamefont {Barras},
  \citenamefont {Roch}, \citenamefont {Brener},\ and\ \citenamefont
  {Bouchbinder}}]{pomyalov2023dynamics}%
  \BibitemOpen
  \bibfield  {author} {\bibinfo {author} {\bibfnamefont {A.}~\bibnamefont
  {Pomyalov}}, \bibinfo {author} {\bibfnamefont {F.}~\bibnamefont {Barras}},
  \bibinfo {author} {\bibfnamefont {T.}~\bibnamefont {Roch}}, \bibinfo {author}
  {\bibfnamefont {E.~A.}\ \bibnamefont {Brener}},\ and\ \bibinfo {author}
  {\bibfnamefont {E.}~\bibnamefont {Bouchbinder}},\ }\bibfield  {title}
  {\bibinfo {title} {The dynamics of unsteady frictional slip pulses},\ }\href
  {https://doi.org/10.1073/pnas.2309374120} {\bibfield  {journal} {\bibinfo
  {journal} {Proceedings of the National Academy of Sciences}\ }\textbf
  {\bibinfo {volume} {120}},\ \bibinfo {pages} {e2309374120} (\bibinfo {year}
  {2023}{\natexlab{b}})}\BibitemShut {NoStop}%
\bibitem [{\citenamefont {Pomyalov}\ and\ \citenamefont
  {Bouchbinder}(2024)}]{pomyalov2024}%
  \BibitemOpen
  \bibfield  {author} {\bibinfo {author} {\bibfnamefont {A.}~\bibnamefont
  {Pomyalov}}\ and\ \bibinfo {author} {\bibfnamefont {E.}~\bibnamefont
  {Bouchbinder}},\ }\bibfield  {title} {\bibinfo {title} {Unsteady slip pulses
  under spatially-varying prestress},\ }\href
  {https://doi.org/https://doi.org/10.1016/j.epsl.2024.119111} {\bibfield
  {journal} {\bibinfo  {journal} {Earth and Planetary Science Letters}\
  }\textbf {\bibinfo {volume} {648}},\ \bibinfo {pages} {119111} (\bibinfo
  {year} {2024})}\BibitemShut {NoStop}%
\bibitem [{\citenamefont {Gabrieli}\ and\ \citenamefont
  {Tal}(2025)}]{gabrieli2025lab}%
  \BibitemOpen
  \bibfield  {author} {\bibinfo {author} {\bibfnamefont {T.}~\bibnamefont
  {Gabrieli}}\ and\ \bibinfo {author} {\bibfnamefont {Y.}~\bibnamefont {Tal}},\
  }\bibfield  {title} {\bibinfo {title} {Lab earthquakes reveal a wide range of
  rupture behaviors controlled by fault bends},\ }\href
  {https://doi.org/10.1073/pnas.2425471122} {\bibfield  {journal} {\bibinfo
  {journal} {Proceedings of the National Academy of Sciences}\ }\textbf
  {\bibinfo {volume} {122}},\ \bibinfo {pages} {e2425471122} (\bibinfo {year}
  {2025})}\BibitemShut {NoStop}%
\bibitem [{\citenamefont {Bouchbinder}(2026)}]{bouchbinder2026equation}%
  \BibitemOpen
  \bibfield  {author} {\bibinfo {author} {\bibfnamefont {E.}~\bibnamefont
  {Bouchbinder}},\ }\bibfield  {title} {\bibinfo {title} {An equation of motion
  for unsteady frictional slip pulses},\ }\href@noop {} {\bibfield  {journal}
  {\bibinfo  {journal} {Earth and Planetary Science Letters}\ }\textbf
  {\bibinfo {volume} {678}},\ \bibinfo {pages} {119831} (\bibinfo {year}
  {2026})}\BibitemShut {NoStop}%
\bibitem [{\citenamefont {Dieterich}(1972)}]{Dieterich1972}%
  \BibitemOpen
  \bibfield  {author} {\bibinfo {author} {\bibfnamefont {J.~H.}\ \bibnamefont
  {Dieterich}},\ }\bibfield  {title} {\bibinfo {title} {{Time-dependent
  friction in rocks}},\ }\href {https://doi.org/10.1029/JB077i020p03690}
  {\bibfield  {journal} {\bibinfo  {journal} {J. Geophys. Res.}\ }\textbf
  {\bibinfo {volume} {77}},\ \bibinfo {pages} {3690} (\bibinfo {year}
  {1972})}\BibitemShut {NoStop}%
\bibitem [{\citenamefont {Dieterich}(1979)}]{Dieterich1979}%
  \BibitemOpen
  \bibfield  {author} {\bibinfo {author} {\bibfnamefont {J.~H.}\ \bibnamefont
  {Dieterich}},\ }\bibfield  {title} {\bibinfo {title} {{Modeling of rock
  friction: 1. Experimental results and constitutive equations}},\ }\href
  {https://doi.org/10.1029/JB084iB05p02161} {\bibfield  {journal} {\bibinfo
  {journal} {J. Geophys. Res. Solid Earth}\ }\textbf {\bibinfo {volume} {84}},\
  \bibinfo {pages} {2161} (\bibinfo {year} {1979})}\BibitemShut {NoStop}%
\bibitem [{\citenamefont {Ruina}(1983)}]{Ruina1983}%
  \BibitemOpen
  \bibfield  {author} {\bibinfo {author} {\bibfnamefont {A.~L.}\ \bibnamefont
  {Ruina}},\ }\bibfield  {title} {\bibinfo {title} {{Slip instability and state
  variable friction laws}},\ }\href {https://doi.org/10.1029/JB088iB12p10359}
  {\bibfield  {journal} {\bibinfo  {journal} {J. Geophys. Res.}\ }\textbf
  {\bibinfo {volume} {88}},\ \bibinfo {pages} {10359} (\bibinfo {year}
  {1983})}\BibitemShut {NoStop}%
\bibitem [{\citenamefont {Marone}(1998)}]{Marone1998a}%
  \BibitemOpen
  \bibfield  {author} {\bibinfo {author} {\bibfnamefont {C.}~\bibnamefont
  {Marone}},\ }\bibfield  {title} {\bibinfo {title} {{Laboratoty-derived
  friction laws and their application to seismic faulting}},\ }\href
  {https://doi.org/10.1146/annurev.earth.26.1.643} {\bibfield  {journal}
  {\bibinfo  {journal} {Annu. Rev. Earth Planet. Sci.}\ }\textbf {\bibinfo
  {volume} {26}},\ \bibinfo {pages} {643} (\bibinfo {year} {1998})}\BibitemShut
  {NoStop}%
\bibitem [{\citenamefont {Nakatani}(2001)}]{Nakatani2001}%
  \BibitemOpen
  \bibfield  {author} {\bibinfo {author} {\bibfnamefont {M.}~\bibnamefont
  {Nakatani}},\ }\bibfield  {title} {\bibinfo {title} {{Conceptual and physical
  clarification of rate and state friction: Frictional sliding as a thermally
  activated rheology}},\ }\href {https://doi.org/10.1029/2000JB900453}
  {\bibfield  {journal} {\bibinfo  {journal} {J. Geophys. Res. Solid Earth}\
  }\textbf {\bibinfo {volume} {106}},\ \bibinfo {pages} {13347} (\bibinfo
  {year} {2001})}\BibitemShut {NoStop}%
\bibitem [{\citenamefont {Baumberger}\ and\ \citenamefont
  {Caroli}(2006)}]{Baumberger2006Solid}%
  \BibitemOpen
  \bibfield  {author} {\bibinfo {author} {\bibfnamefont {T.}~\bibnamefont
  {Baumberger}}\ and\ \bibinfo {author} {\bibfnamefont {C.}~\bibnamefont
  {Caroli}},\ }\bibfield  {title} {\bibinfo {title} {Solid friction from
  stick-slip down to pinning and aging},\ }\href
  {https://doi.org/https://doi.org/10.1080/00018730600732186} {\bibfield
  {journal} {\bibinfo  {journal} {Advances in Physics}\ }\textbf {\bibinfo
  {volume} {55}},\ \bibinfo {pages} {279} (\bibinfo {year} {2006})}\BibitemShut
  {NoStop}%
\bibitem [{\citenamefont {Ben-David}\ \emph {et~al.}(2010)\citenamefont
  {Ben-David}, \citenamefont {Rubinstein},\ and\ \citenamefont
  {Fineberg}}]{Ben-David2010}%
  \BibitemOpen
  \bibfield  {author} {\bibinfo {author} {\bibfnamefont {O.}~\bibnamefont
  {Ben-David}}, \bibinfo {author} {\bibfnamefont {S.~M.}\ \bibnamefont
  {Rubinstein}},\ and\ \bibinfo {author} {\bibfnamefont {J.}~\bibnamefont
  {Fineberg}},\ }\bibfield  {title} {\bibinfo {title} {{Slip-stick and the
  evolution of frictional strength}},\ }\href
  {https://doi.org/10.1038/nature08676} {\bibfield  {journal} {\bibinfo
  {journal} {Nature}\ }\textbf {\bibinfo {volume} {463}},\ \bibinfo {pages}
  {76} (\bibinfo {year} {2010})}\BibitemShut {NoStop}%
\bibitem [{\citenamefont {Weng}\ and\ \citenamefont
  {Ampuero}(2019)}]{weng2019dynamics}%
  \BibitemOpen
  \bibfield  {author} {\bibinfo {author} {\bibfnamefont {H.}~\bibnamefont
  {Weng}}\ and\ \bibinfo {author} {\bibfnamefont {J.-P.}\ \bibnamefont
  {Ampuero}},\ }\bibfield  {title} {\bibinfo {title} {The dynamics of elongated
  earthquake ruptures},\ }\href {https://doi.org/10.1029/2019JB017684}
  {\bibfield  {journal} {\bibinfo  {journal} {Journal of Geophysical Research:
  Solid Earth}\ }\textbf {\bibinfo {volume} {124}},\ \bibinfo {pages} {8584}
  (\bibinfo {year} {2019})}\BibitemShut {NoStop}%
\bibitem [{\citenamefont {Shlomai}\ and\ \citenamefont
  {Fineberg}(2016)}]{Shlomai2016}%
  \BibitemOpen
  \bibfield  {author} {\bibinfo {author} {\bibfnamefont {H.}~\bibnamefont
  {Shlomai}}\ and\ \bibinfo {author} {\bibfnamefont {J.}~\bibnamefont
  {Fineberg}},\ }\bibfield  {title} {\bibinfo {title} {{The structure of
  slip-pulses and supershear ruptures driving slip in bimaterial friction}},\
  }\href {https://doi.org/10.1038/ncomms11787} {\bibfield  {journal} {\bibinfo
  {journal} {Nat. Commun.}\ }\textbf {\bibinfo {volume} {7}},\ \bibinfo {pages}
  {11787} (\bibinfo {year} {2016})}\BibitemShut {NoStop}%
\bibitem [{\citenamefont {Poles}\ \emph {et~al.}(2024)\citenamefont {Poles},
  \citenamefont {Shi},\ and\ \citenamefont {Fineberg}}]{poles2024slip}%
  \BibitemOpen
  \bibfield  {author} {\bibinfo {author} {\bibfnamefont {Y.}~\bibnamefont
  {Poles}}, \bibinfo {author} {\bibfnamefont {S.}~\bibnamefont {Shi}},\ and\
  \bibinfo {author} {\bibfnamefont {J.}~\bibnamefont {Fineberg}},\ }\bibfield
  {title} {\bibinfo {title} {Slip-pulses drive frictional motion of dissimilar
  materials: Universality, dynamics, and evolution},\ }\href
  {https://doi.org/10.1073/pnas.2411959121} {\bibfield  {journal} {\bibinfo
  {journal} {Proceedings of the National Academy of Sciences}\ }\textbf
  {\bibinfo {volume} {121}},\ \bibinfo {pages} {e2411959121} (\bibinfo {year}
  {2024})}\BibitemShut {NoStop}%
\bibitem [{\citenamefont {Das}(1980)}]{das1980numerical}%
  \BibitemOpen
  \bibfield  {author} {\bibinfo {author} {\bibfnamefont {S.}~\bibnamefont
  {Das}},\ }\bibfield  {title} {\bibinfo {title} {A numerical method for
  determination of source time functions for general three-dimensional rupture
  propagation},\ }\href {https://doi.org/10.1111/j.1365-246X.1980.tb02593.x}
  {\bibfield  {journal} {\bibinfo  {journal} {Geophysical Journal
  International}\ }\textbf {\bibinfo {volume} {62}},\ \bibinfo {pages} {591}
  (\bibinfo {year} {1980})}\BibitemShut {NoStop}%
\bibitem [{\citenamefont {Tullis}\ and\ \citenamefont
  {Weeks}(1986)}]{Tullis1986}%
  \BibitemOpen
  \bibfield  {author} {\bibinfo {author} {\bibfnamefont {T.~E.}\ \bibnamefont
  {Tullis}}\ and\ \bibinfo {author} {\bibfnamefont {J.~D.}\ \bibnamefont
  {Weeks}},\ }\bibfield  {title} {\bibinfo {title} {{Constitutive behavior and
  stability of frictional sliding of granite}},\ }\href
  {https://doi.org/10.1007/BF00877209} {\bibfield  {journal} {\bibinfo
  {journal} {Pure Appl. Geophys.}\ }\textbf {\bibinfo {volume} {124}},\
  \bibinfo {pages} {383} (\bibinfo {year} {1986})}\BibitemShut {NoStop}%
\bibitem [{\citenamefont {Kilgore}\ \emph {et~al.}(1993)\citenamefont
  {Kilgore}, \citenamefont {Blanpied},\ and\ \citenamefont
  {Dieterich}}]{Kilgore1993}%
  \BibitemOpen
  \bibfield  {author} {\bibinfo {author} {\bibfnamefont {B.~D.}\ \bibnamefont
  {Kilgore}}, \bibinfo {author} {\bibfnamefont {M.~L.}\ \bibnamefont
  {Blanpied}},\ and\ \bibinfo {author} {\bibfnamefont {J.~H.}\ \bibnamefont
  {Dieterich}},\ }\bibfield  {title} {\bibinfo {title} {{Velocity dependent
  friction of granite over a wide range of conditions}},\ }\href
  {https://doi.org/10.1029/93GL00368} {\bibfield  {journal} {\bibinfo
  {journal} {Geophys. Res. Lett.}\ }\textbf {\bibinfo {volume} {20}},\ \bibinfo
  {pages} {903} (\bibinfo {year} {1993})}\BibitemShut {NoStop}%
\bibitem [{\citenamefont {Dieterich}\ and\ \citenamefont
  {Kilgore}(1994)}]{Dieterich1994a}%
  \BibitemOpen
  \bibfield  {author} {\bibinfo {author} {\bibfnamefont {J.~H.}\ \bibnamefont
  {Dieterich}}\ and\ \bibinfo {author} {\bibfnamefont {B.~D.}\ \bibnamefont
  {Kilgore}},\ }\bibfield  {title} {\bibinfo {title} {{Direct observation of
  frictional contacts: New insights for state-dependent properties}},\ }\href
  {https://doi.org/10.1007/BF00874332} {\bibfield  {journal} {\bibinfo
  {journal} {Pure Appl. Geophys.}\ }\textbf {\bibinfo {volume} {143}},\
  \bibinfo {pages} {283} (\bibinfo {year} {1994})}\BibitemShut {NoStop}%
\bibitem [{\citenamefont {Rubinstein}\ \emph {et~al.}(2004)\citenamefont
  {Rubinstein}, \citenamefont {Cohen},\ and\ \citenamefont
  {Fineberg}}]{Rubinstein2004}%
  \BibitemOpen
  \bibfield  {author} {\bibinfo {author} {\bibfnamefont {S.~M.}\ \bibnamefont
  {Rubinstein}}, \bibinfo {author} {\bibfnamefont {G.}~\bibnamefont {Cohen}},\
  and\ \bibinfo {author} {\bibfnamefont {J.}~\bibnamefont {Fineberg}},\
  }\bibfield  {title} {\bibinfo {title} {{Detachment fronts and the onset of
  dynamic friction}},\ }\href {https://doi.org/10.1038/nature02830} {\bibfield
  {journal} {\bibinfo  {journal} {Nature}\ }\textbf {\bibinfo {volume} {430}},\
  \bibinfo {pages} {1005} (\bibinfo {year} {2004})}\BibitemShut {NoStop}%
\bibitem [{\citenamefont {Marone}\ \emph {et~al.}(1990)\citenamefont {Marone},
  \citenamefont {Raleigh},\ and\ \citenamefont {Scholz}}]{Marone1990}%
  \BibitemOpen
  \bibfield  {author} {\bibinfo {author} {\bibfnamefont {C.}~\bibnamefont
  {Marone}}, \bibinfo {author} {\bibfnamefont {C.~B.}\ \bibnamefont
  {Raleigh}},\ and\ \bibinfo {author} {\bibfnamefont {C.~H.}\ \bibnamefont
  {Scholz}},\ }\bibfield  {title} {\bibinfo {title} {{Frictional behavior and
  constitutive modeling of simulated fault gouge}},\ }\href
  {https://doi.org/10.1029/JB095iB05p07007} {\bibfield  {journal} {\bibinfo
  {journal} {J. Geophys. Res.}\ }\textbf {\bibinfo {volume} {95}},\ \bibinfo
  {pages} {7007} (\bibinfo {year} {1990})}\BibitemShut {NoStop}%
\bibitem [{\citenamefont {Rice}(2006)}]{Rice2006}%
  \BibitemOpen
  \bibfield  {author} {\bibinfo {author} {\bibfnamefont {J.~R.}\ \bibnamefont
  {Rice}},\ }\bibfield  {title} {\bibinfo {title} {{Heating and weakening of
  faults during earthquake slip}},\ }\href
  {https://doi.org/10.1029/2005JB004006} {\bibfield  {journal} {\bibinfo
  {journal} {J. Geophys. Res. Solid Earth}\ }\textbf {\bibinfo {volume}
  {111}},\ \bibinfo {pages} {B05311} (\bibinfo {year} {2006})}\BibitemShut
  {NoStop}%
\bibitem [{\citenamefont {Viesca}\ and\ \citenamefont
  {Garagash}(2015)}]{Viesca2015}%
  \BibitemOpen
  \bibfield  {author} {\bibinfo {author} {\bibfnamefont {R.~C.}\ \bibnamefont
  {Viesca}}\ and\ \bibinfo {author} {\bibfnamefont {D.~I.}\ \bibnamefont
  {Garagash}},\ }\bibfield  {title} {\bibinfo {title} {{Ubiquitous weakening of
  faults due to thermal pressurization}},\ }\href
  {https://doi.org/10.1038/ngeo2554} {\bibfield  {journal} {\bibinfo  {journal}
  {Nat. Geosci.}\ }\textbf {\bibinfo {volume} {8}},\ \bibinfo {pages} {875}
  (\bibinfo {year} {2015})}\BibitemShut {NoStop}%
\bibitem [{\citenamefont {Rubino}\ \emph {et~al.}(2017)\citenamefont {Rubino},
  \citenamefont {Rosakis},\ and\ \citenamefont {Lapusta}}]{Rubino2017}%
  \BibitemOpen
  \bibfield  {author} {\bibinfo {author} {\bibfnamefont {V.}~\bibnamefont
  {Rubino}}, \bibinfo {author} {\bibfnamefont {A.~J.}\ \bibnamefont
  {Rosakis}},\ and\ \bibinfo {author} {\bibfnamefont {N.}~\bibnamefont
  {Lapusta}},\ }\bibfield  {title} {\bibinfo {title} {{Understanding dynamic
  friction through spontaneously evolving laboratory earthquakes}},\ }\href
  {https://doi.org/10.1038/ncomms15991} {\bibfield  {journal} {\bibinfo
  {journal} {Nat. Commun.}\ }\textbf {\bibinfo {volume} {8}},\ \bibinfo {pages}
  {15991} (\bibinfo {year} {2017})}\BibitemShut {NoStop}%
\bibitem [{\citenamefont {Rubino}\ \emph {et~al.}(2019)\citenamefont {Rubino},
  \citenamefont {Rosakis},\ and\ \citenamefont {Lapusta}}]{rubino2019full}%
  \BibitemOpen
  \bibfield  {author} {\bibinfo {author} {\bibfnamefont {V.}~\bibnamefont
  {Rubino}}, \bibinfo {author} {\bibfnamefont {A.}~\bibnamefont {Rosakis}},\
  and\ \bibinfo {author} {\bibfnamefont {N.}~\bibnamefont {Lapusta}},\
  }\bibfield  {title} {\bibinfo {title} {Full-field ultrahigh-speed
  quantification of dynamic shear ruptures using digital image correlation},\
  }\href {https://doi.org/10.1007/s11340-019-00501-7} {\bibfield  {journal}
  {\bibinfo  {journal} {Experimental Mechanics}\ }\textbf {\bibinfo {volume}
  {59}},\ \bibinfo {pages} {551} (\bibinfo {year} {2019})}\BibitemShut
  {NoStop}%
\bibitem [{\citenamefont {Tal}\ \emph {et~al.}(2020)\citenamefont {Tal},
  \citenamefont {Rubino}, \citenamefont {Rosakis},\ and\ \citenamefont
  {Lapusta}}]{tal2020illuminating}%
  \BibitemOpen
  \bibfield  {author} {\bibinfo {author} {\bibfnamefont {Y.}~\bibnamefont
  {Tal}}, \bibinfo {author} {\bibfnamefont {V.}~\bibnamefont {Rubino}},
  \bibinfo {author} {\bibfnamefont {A.~J.}\ \bibnamefont {Rosakis}},\ and\
  \bibinfo {author} {\bibfnamefont {N.}~\bibnamefont {Lapusta}},\ }\bibfield
  {title} {\bibinfo {title} {Illuminating the physics of dynamic friction
  through laboratory earthquakes on thrust faults},\ }\href
  {https://doi.org/10.1073/pnas.2004590117} {\bibfield  {journal} {\bibinfo
  {journal} {Proceedings of the National Academy of Sciences}\ }\textbf
  {\bibinfo {volume} {117}},\ \bibinfo {pages} {21095} (\bibinfo {year}
  {2020})}\BibitemShut {NoStop}%
\bibitem [{\citenamefont {Rubino}\ \emph {et~al.}(2020)\citenamefont {Rubino},
  \citenamefont {Rosakis},\ and\ \citenamefont
  {Lapusta}}]{rubino2020spatiotemporal}%
  \BibitemOpen
  \bibfield  {author} {\bibinfo {author} {\bibfnamefont {V.}~\bibnamefont
  {Rubino}}, \bibinfo {author} {\bibfnamefont {A.}~\bibnamefont {Rosakis}},\
  and\ \bibinfo {author} {\bibfnamefont {N.}~\bibnamefont {Lapusta}},\
  }\bibfield  {title} {\bibinfo {title} {Spatiotemporal properties of
  sub-rayleigh and supershear ruptures inferred from full-field dynamic imaging
  of laboratory experiments},\ }\href {https://doi.org/10.1029/2019JB018922}
  {\bibfield  {journal} {\bibinfo  {journal} {Journal of Geophysical Research:
  Solid Earth}\ }\textbf {\bibinfo {volume} {125}},\ \bibinfo {pages}
  {e2019JB018922} (\bibinfo {year} {2020})}\BibitemShut {NoStop}%
\bibitem [{\citenamefont {Persson}(1998)}]{Persson1998}%
  \BibitemOpen
  \bibfield  {author} {\bibinfo {author} {\bibfnamefont {B.~N.~J.}\
  \bibnamefont {Persson}},\ }\href {https://doi.org/10.1007/978-3-662-04283-0}
  {\emph {\bibinfo {title} {{Sliding friction: physical principles and
  applications}}}}\ (\bibinfo  {publisher} {Springer Sciense {\&} Buisness
  Media},\ \bibinfo {year} {1998})\BibitemShut {NoStop}%
\bibitem [{\citenamefont {Roch}\ \emph {et~al.}(2024)\citenamefont {Roch},
  \citenamefont {Brener}, \citenamefont {Molinari},\ and\ \citenamefont
  {Bouchbinder}}]{roch2024finite}%
  \BibitemOpen
  \bibfield  {author} {\bibinfo {author} {\bibfnamefont {T.}~\bibnamefont
  {Roch}}, \bibinfo {author} {\bibfnamefont {E.~A.}\ \bibnamefont {Brener}},
  \bibinfo {author} {\bibfnamefont {J.-F.}\ \bibnamefont {Molinari}},\ and\
  \bibinfo {author} {\bibfnamefont {E.}~\bibnamefont {Bouchbinder}},\
  }\bibfield  {title} {\bibinfo {title} {A finite geometry, inertia assisted
  coarsening-to-complexity transition in homogeneous frictional systems},\
  }\href {https://doi.org/10.1016/j.jmps.2024.105706} {\bibfield  {journal}
  {\bibinfo  {journal} {Journal of the Mechanics and Physics of Solids}\
  }\textbf {\bibinfo {volume} {190}},\ \bibinfo {pages} {105706} (\bibinfo
  {year} {2024})}\BibitemShut {NoStop}%
\bibitem [{\citenamefont {Crooks}\ \emph {et~al.}(2013)\citenamefont {Crooks},
  \citenamefont {Marsh}, \citenamefont {Turchetta}, \citenamefont {Taylor},
  \citenamefont {Chan}, \citenamefont {Lahav},\ and\ \citenamefont
  {Fenigstein}}]{Crooks2013}%
  \BibitemOpen
  \bibfield  {author} {\bibinfo {author} {\bibfnamefont {J.}~\bibnamefont
  {Crooks}}, \bibinfo {author} {\bibfnamefont {B.}~\bibnamefont {Marsh}},
  \bibinfo {author} {\bibfnamefont {R.}~\bibnamefont {Turchetta}}, \bibinfo
  {author} {\bibfnamefont {K.}~\bibnamefont {Taylor}}, \bibinfo {author}
  {\bibfnamefont {W.}~\bibnamefont {Chan}}, \bibinfo {author} {\bibfnamefont
  {A.}~\bibnamefont {Lahav}},\ and\ \bibinfo {author} {\bibfnamefont
  {A.}~\bibnamefont {Fenigstein}},\ }\bibfield  {title} {\bibinfo {title}
  {{Kirana: a solid-state megapixel uCMOS image sensor for ultrahigh speed
  imaging}},\ }\href {https://doi.org/10.1117/12.2011762} {\bibfield  {journal}
  {\bibinfo  {journal} {Sensors, Cameras, and Systems for Industrial and
  Scientific Applications XIV}\ }\textbf {\bibinfo {volume} {8659}},\ \bibinfo
  {pages} {865903} (\bibinfo {year} {2013})}\BibitemShut {NoStop}%
\bibitem [{\citenamefont {Sutton}\ \emph {et~al.}(2009)\citenamefont {Sutton},
  \citenamefont {Orteu},\ and\ \citenamefont {Schreirer}}]{Sutton2009}%
  \BibitemOpen
  \bibfield  {author} {\bibinfo {author} {\bibfnamefont {M.~A.}\ \bibnamefont
  {Sutton}}, \bibinfo {author} {\bibfnamefont {J.~J.}\ \bibnamefont {Orteu}},\
  and\ \bibinfo {author} {\bibfnamefont {H.}~\bibnamefont {Schreirer}},\ }\href
  {https://doi.org/10.1007/978-0-387-78747-3} {\emph {\bibinfo {title} {{Image
  correlation for shape, motion and deformation measurements: basic concepts,
  theory and applications}}}}\ (\bibinfo  {publisher} {Springer Science \&
  Business Media},\ \bibinfo {year} {2009})\BibitemShut {NoStop}%
\bibitem [{\citenamefont {Buades}\ \emph {et~al.}(2011)\citenamefont {Buades},
  \citenamefont {Coll},\ and\ \citenamefont {Morel}}]{buades2011non}%
  \BibitemOpen
  \bibfield  {author} {\bibinfo {author} {\bibfnamefont {A.}~\bibnamefont
  {Buades}}, \bibinfo {author} {\bibfnamefont {B.}~\bibnamefont {Coll}},\ and\
  \bibinfo {author} {\bibfnamefont {J.-M.}\ \bibnamefont {Morel}},\ }\bibfield
  {title} {\bibinfo {title} {Non-local means denoising},\ }\href
  {https://doi.org/10.5201/ipol.2011.bcm_nlm} {\bibfield  {journal} {\bibinfo
  {journal} {Image processing on line}\ }\textbf {\bibinfo {volume} {1}},\
  \bibinfo {pages} {208} (\bibinfo {year} {2011})}\BibitemShut {NoStop}%
\bibitem [{\citenamefont {Buades}\ \emph {et~al.}(2008)\citenamefont {Buades},
  \citenamefont {Coll},\ and\ \citenamefont {Morel}}]{buades2008nonlocal}%
  \BibitemOpen
  \bibfield  {author} {\bibinfo {author} {\bibfnamefont {A.}~\bibnamefont
  {Buades}}, \bibinfo {author} {\bibfnamefont {B.}~\bibnamefont {Coll}},\ and\
  \bibinfo {author} {\bibfnamefont {J.-M.}\ \bibnamefont {Morel}},\ }\bibfield
  {title} {\bibinfo {title} {Nonlocal image and movie denoising},\ }\href
  {https://doi.org/10.1007/s11263-007-0052-1} {\bibfield  {journal} {\bibinfo
  {journal} {International journal of computer vision}\ }\textbf {\bibinfo
  {volume} {76}},\ \bibinfo {pages} {123} (\bibinfo {year} {2008})}\BibitemShut
  {NoStop}%
\bibitem [{\citenamefont {Duran~Vergara}\ \emph {et~al.}(2024)\citenamefont
  {Duran~Vergara}, \citenamefont {Liebold},\ and\ \citenamefont
  {Maas}}]{duran2024ultrahigh}%
  \BibitemOpen
  \bibfield  {author} {\bibinfo {author} {\bibfnamefont {L.~C.}\ \bibnamefont
  {Duran~Vergara}}, \bibinfo {author} {\bibfnamefont {F.}~\bibnamefont
  {Liebold}},\ and\ \bibinfo {author} {\bibfnamefont {H.-G.}\ \bibnamefont
  {Maas}},\ }\bibfield  {title} {\bibinfo {title} {Ultrahigh-speed imaging for
  high-impact concrete deformation analysis in pre-and post-cracking stages},\
  }\href {https://doi.org/10.1364/ao.506701} {\bibfield  {journal} {\bibinfo
  {journal} {Applied Optics}\ }\textbf {\bibinfo {volume} {63}},\ \bibinfo
  {pages} {467} (\bibinfo {year} {2024})}\BibitemShut {NoStop}%
\bibitem [{\citenamefont {Quino}\ \emph {et~al.}(2021)\citenamefont {Quino},
  \citenamefont {Chen}, \citenamefont {Ramakrishnan}, \citenamefont
  {Mart{\'\i}nez-Hergueta}, \citenamefont {Zumpano}, \citenamefont
  {Pellegrino},\ and\ \citenamefont {Petrinic}}]{quino2021speckle}%
  \BibitemOpen
  \bibfield  {author} {\bibinfo {author} {\bibfnamefont {G.}~\bibnamefont
  {Quino}}, \bibinfo {author} {\bibfnamefont {Y.}~\bibnamefont {Chen}},
  \bibinfo {author} {\bibfnamefont {K.~R.}\ \bibnamefont {Ramakrishnan}},
  \bibinfo {author} {\bibfnamefont {F.}~\bibnamefont {Mart{\'\i}nez-Hergueta}},
  \bibinfo {author} {\bibfnamefont {G.}~\bibnamefont {Zumpano}}, \bibinfo
  {author} {\bibfnamefont {A.}~\bibnamefont {Pellegrino}},\ and\ \bibinfo
  {author} {\bibfnamefont {N.}~\bibnamefont {Petrinic}},\ }\bibfield  {title}
  {\bibinfo {title} {Speckle patterns for dic in challenging scenarios: rapid
  application and impact endurance},\ }\href
  {https://doi.org/10.1088/1361-6501/abaae8} {\bibfield  {journal} {\bibinfo
  {journal} {Measurement Science and Technology}\ }\textbf {\bibinfo {volume}
  {32}},\ \bibinfo {pages} {015203} (\bibinfo {year} {2021})}\BibitemShut
  {NoStop}%
\bibitem [{\citenamefont {Lukic}(2018)}]{Lukic2018}%
  \BibitemOpen
  \bibfield  {author} {\bibinfo {author} {\bibfnamefont {B.}~\bibnamefont
  {Lukic}},\ }\emph {\bibinfo {title} {Development of a full-field measuring
  technique for the characterisation of the dynamic tensile behaviour of
  concrete}},\ \href
  {https://doi.org/10.70675/4c279273z1743z43cez9f9ez372ff10b2196} {Ph.D.
  thesis},\ \bibinfo  {school} {Universit\'e Grenoble Alpes} (\bibinfo {year}
  {2018})\BibitemShut {NoStop}%
\bibitem [{\citenamefont {Gabrieli}(2026)}]{Gabrieli2026}%
  \BibitemOpen
  \bibfield  {author} {\bibinfo {author} {\bibfnamefont {T.}~\bibnamefont
  {Gabrieli}},\ }\emph {\bibinfo {title} {The Effects of Fault Roughness on the
  Dynamics of Earthquake Ruptures}},\ \href@noop {} {Ph.D. thesis},\ \bibinfo
  {school} {Ben-Gurion University of the Negev} (\bibinfo {year}
  {2026})\BibitemShut {NoStop}%
\end{thebibliography}

%

\end{document}